\documentclass[10pt,conference]{IEEEtran}

\usepackage{xspace}
\usepackage{tcolorbox}
\usepackage{enumitem}
\usepackage[ruled]{algorithm2e}
\usepackage[numbers,sort]{natbib}
\usepackage[colorlinks=true,citecolor=blue,linkcolor=blue,urlcolor=blue,bookmarks=false]{hyperref}
\usepackage{booktabs} 
\usepackage{multirow}
\usepackage{graphicx}
\usepackage{tabularx}
\newcolumntype{C}[1]{>{\centering\arraybackslash}p{#1}}
\usepackage{colortbl}
\usepackage{graphicx}  
\usepackage{float}  
\usepackage{subfig}
\usepackage{makecell} 
\usepackage{color}
\usepackage{arydshln} 
\usepackage{caption}
\usepackage{wrapfig}
\usepackage{colortbl}
\usepackage{amsthm} 
\usepackage{amsmath}
\usepackage{adjustbox}
\usepackage{threeparttable}
\usepackage{stfloats}
\usepackage{pifont}
\usepackage{xurl}
\usepackage{amsfonts}
\usepackage{dsfont}

\newcommand{\tool}{\textsc{CoSafe}\xspace}
\newcommand{\todo}[1]{\textcolor{black}{#1}}
\newcommand{\CR}[1]{\textcolor{black}{#1}}
\newcommand{\ie}[0]{\textit{i.e.,}\xspace}
\newcommand{\eg}[0]{\textit{e.g.,}\xspace}

\definecolor{Gray}{gray}{0.9}
\definecolor{LightCyan}{rgb}{0.88,1,1}
\usepackage{microtype}
\def\BibTeX{{\rm B\kern-.05em{\sc i\kern-.025em b}\kern-.08em
    T\kern-.1667em\lower.7ex\hbox{E}\kern-.125emX}}

\begin{document}

\title{\tool: Cooperative Runtime Safety Assurance for Mixed-Traffic Autonomous Driving}

\author{
\IEEEauthorblockN{
  Dingji Wang\IEEEauthorrefmark{1}, You Lu\IEEEauthorrefmark{1}, 
  Haowen Jiang\IEEEauthorrefmark{1},
  Yanggang Chen\IEEEauthorrefmark{1},
  Yuyang Bi\IEEEauthorrefmark{1}, \\
  Bihuan Chen\IEEEauthorrefmark{1}\IEEEauthorrefmark{3}, 
  Shenglin Bao\IEEEauthorrefmark{2}, 
  Xin Peng\IEEEauthorrefmark{1}
}

\IEEEauthorblockA{\IEEEauthorrefmark{1}College of Computer Science and Artificial Intelligence, Fudan University, Shanghai, China}
\IEEEauthorblockA{\IEEEauthorrefmark{2}East China Institute of Computing Technology, Shanghai, China}
\IEEEauthorblockA{\IEEEauthorrefmark{3}Corresponding Author}
}

\maketitle

\begin{abstract}
With the rapid development of autonomous driving systems (ADSes), real-world roads are evolving into mixed-traffic environments where autonomous vehicles (AVs) and \todo{other traffic participants (\eg human-driven vehicles (HDVs), cyclists, and pedestrians)} coexist and interact with each other.~However,~current ADSes still incur various safety violations in mixed~traffic scenarios, raising a strong demand for runtime safety assurance frameworks that can continuously monitor hazards and prevent~potential~violations. 
Existing ego-centric assurance approaches are insufficient for mixed traffic, where violations
often arise from locally~reasonable yet jointly incompatible decisions among multiple vehicles and cannot be reliably resolved through independent takeover.

To bridge this gap, we propose \tool, a modular~and~ADS-agnostic cooperative runtime safety assurance framework for mixed-traffic autonomous driving to extend runtime safety~assurance from ego-centric takeover to scene-level cooperative~intervention. \tool leverages vehicle-to-everything (V2X) communication to enable roadside units (RSUs) to monitor scene-level hazards and jointly mitigate hazards involving multiple vehicles. We integrate \tool with seven state-of-the-art ADSes,~and~our evaluation on the InterDrive benchmark demonstrates that~\tool improves the driving score of AVs by \todo{70.03\%} on average, reduces safety violations by \todo{87.80\%}, and accurately identifies unsafe AVs with an $\text{F}_3$ score of \todo{0.930}, while satisfying real-time constraints.

\end{abstract}

\section{Introduction}

With the rapid development of ADSes, real-world roads~are~increasingly evolving into mixed-traffic environments where~AVs (\ie vehicles equipped~with ADSes) and \todo{other traffic participants (\eg HDVs, cyclists, and pedestrians)} coexist~and~interact~\cite{wang2025knowledge}. %
However, the performance of AVs in open-road~mixed traffic remains far from satisfactory. Scenario-based testing approaches \cite{cheng2023behavexplor, zhang2023building, lu2024advfuzz} have indicated that even advanced ADSes may still incur safety violations in mixed-traffic environments. Besides,~a 2024 U.S. Department of Transportation analysis of California ADS disengagement reports~\cite{khan2024understanding} found that 69\%~of~disengagements were caused by ADS insufficiency.~These~limitations~underscore the need for runtime safety assurance frameworks to prevent safety violations in mixed-traffic environments.

Runtime safety assurance~\cite{wang2025argus} for ADSes aims to continuously monitor driving hazards, intervene before potential~safety~violations occur, and maintain safe operation~at~runtime.~To~prevent safety violations, several approaches monitor~basic~safety~properties and trigger emergency braking~\cite{kusano2012safety, zhang2024dual} or conservative fallback maneuvers~\cite{chen2022runtime, peng2023rta} before these properties~are~violated. \todo{However, these approaches rely on predetermined emergency interventions rather than adapting to the specific hazardous scenarios encountered. To~intervene~more adaptively,~REDriver~\cite{sun2024redriver} intervenes in the planning process by refining ADS-generated \CR{planned} trajectories to enforce specification compliance, but it requires intrusive modifications to the original ADS stack, and~is~tied~to Apollo~\cite{apollo}. Without modifying the underlying ADSes, \textsc{Argus} \cite{wang2025argus} formulates runtime assurance as a hazard-aware takeover-and-recovery process for the trajectory of one single AV.}

\begin{figure}[!t]
    \centering
    \subfloat[\scriptsize ADS Conflict]{%
        \includegraphics[width=0.31\linewidth]{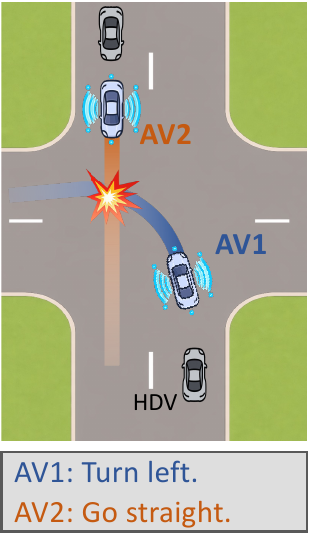}%
        \label{img:motivation:a}}
    \hspace{0.002\linewidth}
    \subfloat[\scriptsize Ego-centric Takeover]{%
        \includegraphics[width=0.31\linewidth]{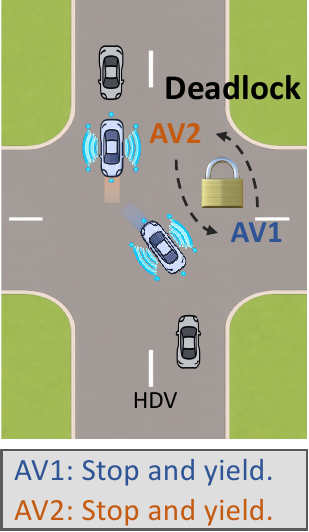}%
        \label{img:motivation:b}}
    \hspace{0.002\linewidth}
    \subfloat[\scriptsize Scene-level Intervention]{%
        \includegraphics[width=0.31\linewidth]{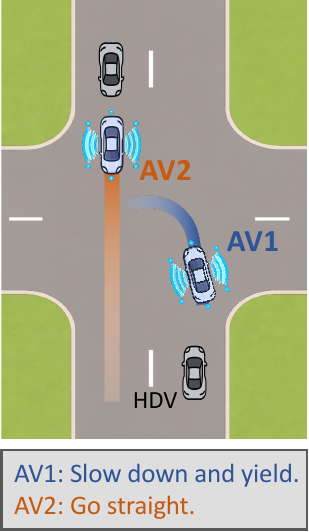}%
        \label{img:motivation:c}}
    \vspace{-1pt}
    \caption{Motivation of Cooperative Runtime Safety Assurance}
    \label{img:motivation}
\end{figure}

However, these ego-centric approaches assess and mitigate hazards on a per-vehicle basis, treating each AV~as~an~independent takeover target rather than viewing mixed traffic as a software-intensive cyber-physical system of interacting AVs, \todo{other traffic participants,} and RSUs~\cite{rajkumar2010cyber}. As shown in Fig.~\ref{img:motivation:a}, two ADS-controlled AVs may generate trajectories that conflict at an unsignalized intersection. Although independently taking over each AV can avoid an immediate collision, the two AVs may enter a circular wait-for chain~because~neither~vehicle~has~\todo{a scene-level view to determine yield priority},~causing~a~deadlock in Fig.~\ref{img:motivation:b}. Mixed-traffic safety violations often arise~from~locally reasonable yet jointly incompatible decisions among~multiple vehicles, which cannot be reliably resolved~via~ego-centric takeover. As shown in Fig.~\ref{img:motivation:c}, resolving this interaction-induced safety violation requires scene-level intervention, \todo{where~one~AV is assigned to proceed while the other slows down and yields.}

\todo{Such intervention, which no single AV can reliably produce on its own, requires a scene-level view across multiple vehicles. V2X communication provides a practical foundation for this view, allowing AVs to exchange state, intent, and agreement-seeking information with RSUs.}
\todo{However, most existing V2X-enabled cooperative driving approaches still~leave~the~driving decisions to individual vehicles, which~may~incur~safety~violations in mixed traffic. Representative work includes shared perception~\cite{lu2024extensible,wang2024deepaccident,liu2025towards,yu2025end,cui2022coopernaut}, decentralized multi-vehicle planning~\cite{zhao2023multi,chen2024communication,zhou2022cooperative}, and LLM-mediated negotiation~\cite{fang2025towards,liu2025colmdriver,gao2025langcoop}.}
In contrast, centralized multi-vehicle planning approaches~\cite{liu2024improved,huang2023parallel,huang2023decentralized} coordinate traffic behavior at the scene level, but they typically assume an AV-dominated fleet with homogeneous ADSes under direct centralized control, a setting at odds with real-world mixed traffic, where AVs run heterogeneous,~vendor-specific ADSes. \CR{Migrating such coordination into a safety assurance layer is non-trivial, as ADSes are black boxes and mixed traffic is only partially controllable (\ie HDVs and pedestrians cannot be overridden, and safe AVs should not be).}
\todo{In summary, existing work lacks V2X-enabled scene-level safety assurance~for~mixed~traffic with heterogeneous AVs. To the best of our knowledge, no existing work monitors scene-level hazards and jointly intervenes on unsafe AVs without modifying their underlying planning logic.}

To bridge this gap, we present \tool, a modular~and~ADS-agnostic cooperative runtime safety assurance framework for mixed-traffic autonomous driving that extends runtime safety assurance from ego-centric takeover to scene-level~cooperative intervention.
Specifically, \tool consists of three RSU-side modules, namely \textit{cooperative perception}, \textit{hazard assessment}, and \textit{joint hazard mitigation}, together with an AV-side \todo{\textit{control switch}} module. Within the coverage area~of~an~RSU,~the~cooperative perception aggregates V2X information from participating AVs with RSU-side observations to construct the~scene-level traffic information. Based on this information,~the~hazard~assessment detects potential driving hazards, and identifies~corresponding unsafe AVs; and the joint hazard mitigation jointly generates safe reference trajectories for unsafe AVs. For each participating AV, the control switch either passes through the original ADS-generated \CR{planned} trajectory if no intervention is needed, or executes the generated safe reference trajectory. %

We conduct large-scale experiments to evaluate the effectiveness and efficiency of \tool in mixed-traffic environments. We integrate \tool with seven state-of-the-art ADSes spanning five families,
and evaluate them on the InterDrive benchmark~\cite{liu2025colmdriver}. Our experiments demonstrate that \tool effectively improves the driving score of AVs by \todo{70.03\%}, while reducing safety violations by \todo{87.80\%} on average, with runtime latency satisfying real-time constraints. In addition, \tool accurately identifies unsafe AVs with an $\text{F}_3$ score of \todo{0.930}; and our ablation and mixed-deployment results demonstrate that each module contributes distinctly and that \tool remains effective when coexisting AVs run heterogeneous ADSes.

The main contributions of our work are as follows.
\begin{itemize}[leftmargin=*]
    \item We formulate cooperative runtime safety assurance as a scene-level pass-through-or-override safety layer over multiple ADSes, shifting the safety assurance paradigm from ego-centric takeover to scene-level cooperative intervention.

    \item We design \tool, a modular cooperative runtime safety assurance framework, which can be integrated with heterogeneous AVs without modifying their underlying ADSes.

    \item We develop a scene-level hazard assessment and joint mitigation mechanism that identifies unsafe AVs and generates coordinated safe reference trajectories to resolve hazards.

    \item We integrate \tool with seven state-of-the-art ADSes,~and conduct experiments on the InterDrive benchmark to demonstrate its effectiveness, efficiency, and generality.
\end{itemize}

\section{Background and Problem Formulation}
\subsection{Autonomous Driving Systems}

ADSes provide the underlying planning logic of AVs, determining how vehicles behave in dynamic traffic environments~\cite{chen2024end}. Existing ADSes are heterogeneous in their internal architectures, including modular ADSes that separate perception, prediction, planning, and control~\cite{autoware}, end-to-end ADSes that directly map sensor observations to trajectories or control commands~\cite{wu2022trajectoryguided,hu2023planning,jiang2023vad}, LLM-driven ADSes that incorporate language-based reasoning~\cite{shao2024lmdrive}, and V2X-enabled cooperative ADSes that exploit shared perception~\cite{liu2025towards} or negotiation information~\cite{liu2025colmdriver}. 
Despite their architectural differences, they share a functional interface that maps traffic observations to intended driving behaviors.

In this work, we abstract an ADS $\mathcal{ADS}_i$ as a black-box mapping from sensor observations $o_i^t$ and auxiliary planning information $a_i^t$ (\eg target point and high-level maneuver command) to an ADS-generated \CR{planned} trajectory $\tau_i^t$ \CR{at each discrete time step $t$}. Since modern ADSes commonly construct bird's-eye-view (BEV) representations for perception, prediction, or planning, we denote the BEV feature available to \tool as $\mathrm{BEV}_i^t$, as defined by Eq.~\ref{eq:ads-abstraction},
\begin{equation}\label{eq:ads-abstraction}
\small
\begin{array}{l}
    \tau_i^t = \mathcal{ADS}_i(o_i^t, a_i^t),\qquad \mathrm{BEV}_i^t = f_{\mathrm{BEV}}(o_i^t),
\end{array}
\end{equation}
where $\tau_i^t$ is represented as desired waypoints with associated speeds, and $\mathrm{BEV}_i^t$ captures the perceived local traffic scene in a BEV representation. $f_{\mathrm{BEV}}$ denotes the encoder used to obtain $\mathrm{BEV}_i^t$ from raw sensor observations. The trajectory is finally translated by a vehicle~controller~\cite{borase2021review} into control commands.

\subsection{Vehicle-to-Everything Communication}

V2X communication enables AVs, RSUs, and other traffic participants to exchange information for cooperative driving. The SAE J3216 standard~\cite{automation2020taxonomy} divides cooperative driving into four classes, \ie status-sharing, intent-sharing, agreement-seeking, and prescriptive cooperation, where prescriptive cooperation allows an authorized entity, \eg a fleet operations~center or an infrastructure unit, \todo{to direct specific actions to selected AVs, with the receiving AVs adhering to the prescribed actions.}

In this work, we consider an RSU-assisted V2X that~instantiates the prescriptive cooperation only on AVs flagged as unsafe. Within the RSU's coverage area, at each time step $t$, each AV $\mathrm{av}_i$, \todo{running its native ADS $\mathcal{ADS}_i$}, uploads its BEV feature $\mathrm{BEV}_i^t$, vehicle state $s_i^t$ (\ie position $p_i^t$, heading $\theta_i^t$, and speed $v_i^t$), and ADS intention (\ie ADS-generated \CR{planned} trajectory $\tau_i^t$ and auxiliary planning information $a_i^t$) to the RSU, and~the RSU returns a safe reference trajectory $\hat{\tau}_i^t$ to each flagged~unsafe~AV.

\begin{figure*}[!t]
    \centering
    \includegraphics[width=0.9\linewidth]{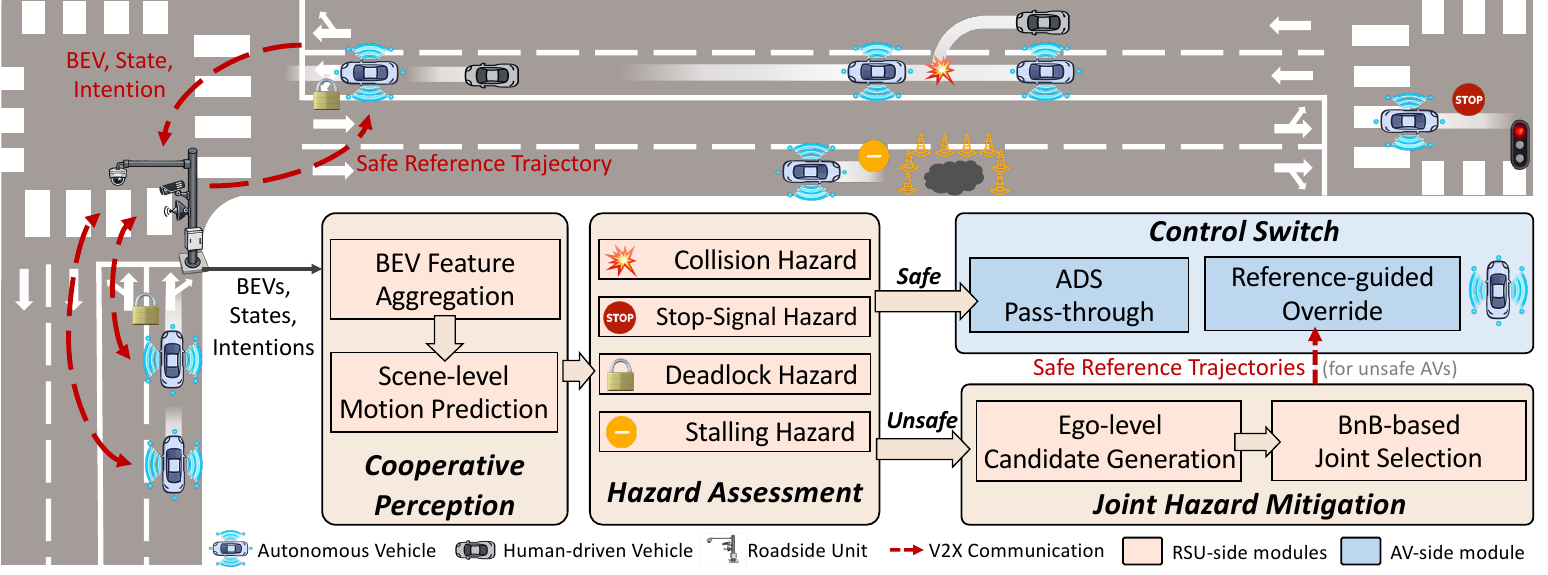}
    \caption{Approach Overview of \tool}
    \label{img:overview}
\end{figure*}

\subsection{Problem Formulation}
\label{sec:problem-formulation}

Runtime safety assurance for ADSes refers to online mechanisms that monitor ADS behavior and intervene when potential safety violations are predicted. In this work, we formulate~runtime safety assurance as a pass-through-or-override safety layer over multiple ADSes. \CR{Formally, the executed trajectory of each AV $\mathrm{av}_i$ is defined by Eq.~\ref{eq:executed-trajectory},}
\begin{equation}\label{eq:executed-trajectory}
\small
\begin{array}{l}
\tilde{\tau}_i^t =
\begin{cases}
\hat{\tau}_i^t, & \mathrm{av}_i \in \mathcal{U}^t, \\
\tau_i^t, & \mathrm{av}_i \notin \mathcal{U}^t,
\end{cases}
\end{array}
\end{equation}
where $\mathcal{U}^t$ denotes the set of unsafe AVs \CR{re-identified at every time step $t$, and thus $\mathrm{av}_i$ is overridden only when it is deemed unsafe.} \CR{Throughout this paper, we distinguish three trajectories of $\mathrm{av}_i$, \ie the ADS-generated planned trajectory $\tau_i^t$ output by its native ADS, the safe reference trajectory $\hat{\tau}_i^t$ generated by the safety layer, and the finally executed trajectory $\tilde{\tau}_i^t$, which equals $\hat{\tau}_i^t$ if $\mathrm{av}_i$ is unsafe and $\tau_i^t$ otherwise.}

Considering a mixed-traffic scene within the coverage area~of an RSU, we denote the set of participating AVs at time step~$t$ as $\mathcal{V}_{\mathrm{AV}}^t$, \todo{and all observed traffic participants (\eg HDVs,~cyclists, pedestrians, as well as AVs) as $\mathcal{TP}^t$}.~Cooperative runtime safety assurance monitors multiple AVs and their scene-level interactions, identifies the unsafe AVs $\mathcal{U}^t \subseteq \mathcal{V}_{\mathrm{AV}}^t$ whose~ADS-generated \CR{planned} trajectories may lead to safety~violations,~and~generates coordinated safe reference trajectories for these AVs. Formally, this problem is formulated by Eq.~\ref{eq:cooperative-rsa},
\begin{equation}\label{eq:cooperative-rsa}
\small
\begin{array}{l}
\left(
\mathcal{U}^t,
\{\hat{\tau}_i^t\}_{\mathrm{av}_i \in \mathcal{U}^t}
\right)
=
\mathcal{R}
\left(
\{(\mathrm{BEV}_i^t, s_i^t, \tau_i^t, a_i^t)\}_{\mathrm{av}_i \in \mathcal{V}_{\mathrm{AV}}^t},
\mathcal{I}^t
\right),
\end{array}
\end{equation}
where $\mathcal{R}$ denotes a cooperative runtime safety assurance~mechanism, and $\mathcal{I}^t = (\mathrm{BEV}_{\mathrm{RSU}}^t, \mathcal{M}_{\mathrm{RSU}}, \mathcal{S}_{\mathrm{RSU}}^t)$ represents RSU-side scene information, including the RSU-side BEV feature $\mathrm{BEV}_{\mathrm{RSU}}^t$, the HD map of the RSU's coverage area $\mathcal{M}_{\mathrm{RSU}}$, and dynamic infrastructure states $\mathcal{S}_{\mathrm{RSU}}^t$ (\eg traffic signal states).

This formulation shifts the safety assurance paradigm from ego-centric takeover to scene-level cooperative intervention. While ego-centric runtime assurance decides whether a single ADS trajectory should be passed through or overridden,~cooperative runtime assurance operates over \todo{multiple AVs}, and produces a coordinated assignment that ensures the resulting scene-level behavior is mutually compatible.

\section{Methodology}

To address the scene-level safety assurance problem defined in Sec.~\ref{sec:problem-formulation}, we propose \tool, a cooperative runtime safety assurance framework for heterogeneous ADSes in mixed-traffic environments, to identify unsafe AVs and generate~coordinated safe reference trajectories at runtime.
The key idea of \tool is to extend runtime safety assurance from ego-centric takeover to scene-level cooperative intervention. Instead of assessing each AV in isolation, \tool jointly evaluates the intentions of multiple AVs under shared scene-level information, identifies the unsafe AVs involved in potential violations, and coordinates their safe reference trajectories. AVs with safe~intentions~continue to execute their ADS-generated \CR{planned} trajectories, while only unsafe AVs are selected to override. \tool is ADS-agnostic, which does not require modifying or retraining the native~ADS, and only relies on the standardized V2X interfaces for state, intention, and scene-feature exchange.

\subsection{Framework Overview}

Fig.~\ref{img:overview} shows the approach overview of \tool. Overall,~it is \todo{designed to preserve a scene-level safety invariant $\Phi_{\mathrm{safe}}^t$,~defined as the absence of four hazards (\ie the collision hazard, the stop-signal hazard, the deadlock hazard and the stalling~hazard)}.
\tool has four modules, including three RSU-side modules, \ie \textit{cooperative perception}, \textit{hazard assessment}, and \textit{joint~hazard mitigation}, and one AV-side module, \ie \textit{control switch}. At each time step $t$, each participating AV $ \mathrm{av}_i$ first executes~its native ADS $\mathcal{ADS}_i$, and obtains a BEV feature $\mathrm{BEV}_i^t$ and~an ADS-generated \CR{planned} trajectory $\tau_i^t$. Then, the AV uploads its BEV feature $\mathrm{BEV}_i^t$, vehicle state $s_i^t$, and ADS intention, including $\tau_i^t$ and $a_i^t$, to the RSU through V2X communication.

On the RSU side, the cooperative perception aggregates the uploaded AV-side information with RSU-side observations to construct the scene-level information of the traffic environment and predict the future motions of observed traffic participants (see Sec.~\ref{sec:coop-perception}). Based on the resulting scene-level information, the hazard assessment evaluates the driving hazards using \todo{a set of hazard oracles} that capture different types of potential safety violations, and identifies the unsafe AV set $\mathcal{U}^t$ that requires intervention (see Sec.~\ref{sec:hazard-assessment}). Then, the joint hazard~mitigation generates candidate trajectories for each unsafe AV, and performs branch-and-bound (BnB)-based joint selection to choose a coordinated safe reference trajectory set $\{\hat{\tau}_i^t\}_{\mathrm{av}_i \in \mathcal{U}^t}$ (see Sec.~\ref{sec:joint-mitigation}). Finally, the RSU sends the generated safe reference trajectories back to the corresponding unsafe AVs.

For an AV that does not receive a safe reference trajectory, the ADS-generated \CR{planned} trajectory $\tau_i^t$ is passed through and executed by the vehicle controller. Receiving a safe reference trajectory indicates that the RSU has identified the AV as unsafe, and \todo{the control switch tracks the assigned trajectory $\hat{\tau}_i^t$ by a reference-guided override to produce control commands that mitigate the potential driving hazards (see Sec.~\ref{sec:control-override}).}

In summary, at each time step $t$, the hazard assessment~identifies AVs whose intended trajectories may violate $\Phi_{\mathrm{safe}}^t$, the joint hazard mitigation selects coordinated safe reference~trajectories from a finite candidate pool, with conservative braking as~a fallback, and the control switch executes the assigned safe reference trajectories on unsafe AVs. \todo{Given accurate cooperative perception and motion prediction, and bounded tracking error,} this closed loop aims to carry the scene from $\Phi_{\mathrm{safe}}^t$~to~$\Phi_{\mathrm{safe}}^{t+1}$.

\subsection{Cooperative Perception}
\label{sec:coop-perception}

To construct the scene-level information, \todo{\ie the states $\bar{s}_k^t$ and future trajectories $\bar{\tau}_k^t$ of all observed traffic participants} together with the surrounding infrastructure states, we implement a cooperative perception module at the RSU side. At each time step $t$, the \textit{BEV feature aggregation} warps and fuses AV-side and RSU-side BEV features into an aggregated BEV feature. Then, the \textit{scene-level motion prediction} predicts the future motions of all observed traffic participants from the aggregated BEV feature, and constructs the scene-level information.

\textbf{BEV Feature Aggregation.}
At each time step $t$, the RSU produces a BEV feature $\mathrm{BEV}_{\mathrm{RSU}}^t$ from its own sensors,~and receives a set of AV-side BEV features $\{\mathrm{BEV}_i^t\}_{\mathrm{av}_i \in \mathcal{V}_{\mathrm{AV}}^t}$~conforming to a shared BEV representation schema (\ie channel dimensions and spatial grid). Since each AV-side BEV feature $\mathrm{BEV}_i^t$ is generated in the \todo{ego coordinate frame} of $\mathrm{av}_i$,~\tool first warps it to the RSU coordinate frame by the \todo{rigid~transformation $T_{i \to \mathrm{RSU}}^t$ from $\mathrm{av}_i$ to the RSU}~\cite{xu2022v2x}, producing the AV-observed features $\mathrm{BEV}_{i \to \mathrm{RSU}}^t$ in the RSU coordinate frame.

Then, \tool fuses all warped BEV features with the RSU-side BEV feature into an aggregated BEV feature by Eq.~\ref{eq:bev-fusion},
\begin{equation}\label{eq:bev-fusion}
  \small
\begin{array}{l}
\mathrm{BEV}_{\mathrm{agg}}^t = \mathcal{F}\!\left(
  \{\mathrm{BEV}_{i \to \mathrm{RSU}}^t\}_{\mathrm{av}_i \in \mathcal{V}_{\mathrm{AV}}^t},\;
  \mathrm{BEV}_{\mathrm{RSU}}^t
\right),
\end{array}
\end{equation}
where $\mathcal{F}$ denotes a V2X fusion model in the RSU coordinate frame, instantiated as a CoBEVT-based transformer fuser~\cite{xu2023cobevt} that jointly attends across the participating AVs and the RSU.

\textbf{Scene-level Motion Prediction.}
Following BEVerse~\cite{zhang2022beverse}, \tool predicts the future motions of all observed traffic participants from the aggregated BEV feature $\mathrm{BEV}_{\mathrm{agg}}^t$ by Eq.~\ref{eq:motion-prediction},
\begin{equation}\label{eq:motion-prediction}
  \small
\begin{array}{l}
\{(\bar{s}_k^t,\, \bar{m}_k^t)\}_{k=1}^{K} = f_{\mathrm{mot}}(\mathrm{BEV}_{\mathrm{agg}}^t),
\end{array}
\end{equation}
where $K$ is the number of observed traffic participants~$\mathcal{TP}^t = \{\mathrm{tp}_k\}_{k=1}^{K}$, and $\bar{s}_k^t$ and $\bar{m}_k^t$ denote the estimated state and~the predicted future motion of the $k$-th participant $\mathrm{tp}_k$. The scene-level motion prediction head $f_{\mathrm{mot}}$ predicts a per-pixel latent map from $\mathrm{BEV}_{\mathrm{agg}}^t$, iteratively warps it forward via predicted flows to generate \todo{$H$-time-step} BEV outputs \todo{(where $H$ is the prediction horizon set to $12$ (\ie $3$ seconds) by default~\cite{wang2024deepaccident}}), and post-processes them via per-instance segmentation and \todo{cross-time-step linking} into the state-motion pairs $\{(\bar{s}_k^t,\, \bar{m}_k^t)\}_{k=1}^{K}$.

Because $f_{\mathrm{mot}}$ operates only on BEV features, its predictions cover all detected traffic participants, but carry~no~V2X~identity. Hence, \tool  matches these predictions with V2X-reported AVs by Hungarian assignment~\cite{kuhn1955hungarian}, where the~matching~cost is computed from the position and heading discrepancies~between detected and reported states. Matched~predictions are considered as AVs, while the remaining predictions~are~treated~as~\todo{other traffic participants}.
\todo{Then,~\tool~constructs the scene-level~information $\mathcal{S}^t = (\{(\bar{s}_k^t,\, \bar{\tau}_k^t)\}_{k=1}^{K},\,\mathcal{M}_{\mathrm{RSU}},\, \mathcal{S}_{\mathrm{RSU}}^t)$, comprising per-participant state-trajectory pairs $\{(\bar{s}_k^t,\, \bar{\tau}_k^t)\}_{k=1}^{K}$, the HD map $\mathcal{M}_{\mathrm{RSU}}$, and the dynamic infrastructure states $\mathcal{S}_{\mathrm{RSU}}^t$. Each pair $(\bar{s}_k^t, \bar{\tau}_k^t)$ is populated according to the participant's V2X identity. Specifically, for matched AVs, \tool sets $\bar{s}_k^t$ to the V2X-reported state $s_i^t$, and $\bar{\tau}_k^t$ to the uploaded ADS-generated \CR{planned} trajectory $\tau_i^t$. For other traffic participants, \tool retains $\bar{s}_k^t$ as $f_{\mathrm{mot}}$'s output, and sets $\bar{\tau}_k^t$ to the predicted motion $\bar{m}_k^t$.}

\subsection{Hazard Assessment}
\label{sec:hazard-assessment}

To identify the unsafe AV set $\mathcal{U}^t$, \tool evaluates~four hazard oracles on the scene-level information $\mathcal{S}^t$, \ie the~\textit{collision hazard}, the \textit{stop-signal hazard}, the \textit{deadlock hazard}, and the \textit{stalling hazard}. \todo{These four hazards cover the most common safety violations in mixed traffic. Among them, the deadlock oracle explicitly detects interaction-induced hazards~across~multiple AVs, complementing the per-vehicle collision and stop-signal oracles in prior ego-centric runtime assurance approaches \cite{sun2024redriver,wang2025argus}.} \tool aggregates the flagged AVs into a \todo{current-time-step} unsafe AV set $\mathcal{U}^t_{\mathrm{cur}}$. To mitigate temporal flickering caused by perception and prediction noise, \tool further augments $\mathcal{U}^t_{\mathrm{cur}}$ with a cooldown set $\mathcal{U}^t_{\mathrm{cool}}$, which retains AVs flagged in recent \todo{time steps}, yielding the unsafe AV set~$\mathcal{U}^t$.

\tool first constructs a per-participant bounding box set $\bar{\mathcal{B}}^t$ over the assessment horizon, defined by Eq.~\ref{eq:box-set},
\begin{equation}\label{eq:box-set}
\small
\begin{array}{l}
\bar{\mathcal{B}}^t = \big\{\, \bar{B}_k^{t'} \;\big|\; \mathrm{tp}_k \in \mathcal{TP}^t,\; t' \in \{t,\ldots,t+H\} \,\big\}, \\
\text{where}\;\; \bar{B}_k^{t'} = \big(\bar{p}_k^{t'},\, \bar{\theta}_k^{t'},\, \bar{v}_k^{t'},\, L_k,\, W_k\big),
\end{array}
\end{equation}
where $H$ is the \todo{prediction horizon from the motion prediction head (Sec.~\ref{sec:coop-perception}), also serving as the assessment horizon}. $\bar{p}_k^{t'}$, $\bar{\theta}_k^{t'}$, and $\bar{v}_k^{t'}$ are the position, heading, and speed of the traffic participant $\mathrm{tp}_k$ at time step $t'$ sampled from the future trajectory $\bar{\tau}_k^t$; and $L_k$ and $W_k$ are its body length and width. Then, we evaluate the following four hazard oracles.

\textbf{Collision Hazard.}
For each $\mathrm{av}_i \in \mathcal{V}_{\mathrm{AV}}^t$, the collision oracle tests whether its predicted bounding boxes $\{\bar{B}_i^{t'}\}_{t' \in \{t,\ldots,t+H\}}$ overlap with those of any other traffic participant at any future \todo{time step} $t'$. 
\CR{Specifically, $\mathrm{IS}(\bar{B}_a^{t'},\, \bar{B}_b^{t'})$ tests whether two bounding boxes intersect after each is inflated by a safety margin $\delta$ to tolerate perception and prediction errors~\cite{wang2025argus}. Following~\cite{gottschalk1996obbtree}, we instantiate it by the separating axis theorem, which determines $\mathrm{IS}(\cdot) = \mathrm{True}$ if and only if the projections of the two boxes overlap on each of their four edge normals.} 
We then define the collision-unsafe AV set $\mathcal{U}^t_{\mathrm{col}}$ by Eq.~\ref{eq:collision-hazard},
\begin{equation}\label{eq:collision-hazard}
\small
\begin{array}{l}
\mathcal{U}^t_{\mathrm{col}} = \big\{\, \mathrm{av}_i \in \mathcal{V}_{\mathrm{AV}}^t \;\big|\; \exists\, \mathrm{tp}_k \in \mathcal{TP}^t \setminus \{\mathrm{av}_i\},\; \\
\hspace{4em} \exists\, t' \in \{t,\ldots,t+H\},\; \mathrm{IS}(\bar{B}_i^{t'},\, \bar{B}_k^{t'}) = \mathrm{True}\,\big\},
\end{array}
\end{equation}
where $\mathrm{tp}_k$ ranges over all perceived traffic participants except $\mathrm{av}_i$, ensuring that each AV is checked for potential collisions with other perceived participants within the assessment horizon.

\textbf{Stop-Signal Hazard.}
A stop-signal violation occurs when~an AV fails to come to a complete stop within the influence~region of an active stop signal (\eg a stop sign or a red light).~We \todo{represent} the influence region of each stop signal as a virtual static bounding box, and \todo{construct} the set of currently active influence boxes $\mathcal{B}_{\mathrm{stop}}^t$ from $\mathcal{M}_{\mathrm{RSU}}$ and $\mathcal{S}_{\mathrm{RSU}}^t$. Then, we define the stop-signal-unsafe AV set $\mathcal{U}^t_{\mathrm{stop}}$ by Eq.~\ref{eq:stop-signal-hazard},
\begin{equation}\label{eq:stop-signal-hazard}
\small
\mathcal{U}^t_{\mathrm{stop}} = \big\{\mathrm{av}_i \in \mathcal{V}_{\mathrm{AV}}^t \big|\exists B \in \mathcal{B}_{\mathrm{stop}}^t, \mathrm{Enter}_i^t(B) \land \neg \mathrm{Stop}_i^t(B) \big\},
\end{equation}
where {\small $\mathrm{Enter}_i^t(B) = \exists\, t' \in \{t,\ldots,t+H\}, \mathrm{IS}(\bar{B}_i^{t'}, B)= \mathrm{True} $}~indicates that $\mathrm{av}_i$'s predicted trajectory enters the active stop signal's influence box $B$, and {\small$\mathrm{Stop}_i^t(B) = \exists\, t' \in \{t,\ldots,t+H\}, \mathrm{IS}(\bar{B}_i^{t'}, B) = \mathrm{True} \land \bar{v}_i^{t'} \leq v_{\mathrm{creep}}$} indicates that $\mathrm{av}_i$ comes to a complete stop while inside $B$, where $v_{\mathrm{creep}}$ is the creep speed threshold, set to $0.1$ m/s by default.

\textbf{Deadlock Hazard.}
Deadlock is an interaction-induced~liveness hazard whose detection requires a scene-level view~of~the wait-for relations across multiple AVs. Inspired by the Coffman conditions~\cite{coffman1971system, cheng2025stclocker}, a deadlock hazard occurs when a set of AVs form a circular wait-for chain such that each AV is blocked by the next AV's currently occupied or planned road region, leaving none able to make forward progress unless at least one changes its intention. To this end, we detect deadlocks in two steps, \ie identifying persistently stalled AVs, and determining whether their wait-for relations form a directed cycle.

We first define a stalling indicator $\mathrm{Stall}_k^t$ that marks whether $\mathrm{tp}_k \in \mathcal{TP}^t$ has been persistently stationary without legitimate cause, as formalized in Eq.~\ref{eq:stall-indicator},
\begin{equation}\label{eq:stall-indicator}
\small
\mathrm{Stall}_k^t = \mathds{1}\!\left[\forall t' \in \mathcal{T}_W,\ v_k^{t'} < v_{\mathrm{creep}} \land \neg\, \mathrm{ValidStop}(\mathrm{tp}_k, t')\right],
\end{equation}
where $\mathds{1}[\cdot]$ is the indicator function, $\mathcal{T}_W=\{t-W+1,\ldots,t\}$~is a \todo{past-time-step} window of length $W$, and \CR{$v_k^{t'}$ is $\mathrm{tp}_k$'s speed at past time step $t'$.} $\mathrm{ValidStop}(\mathrm{tp}_k,t')$ checks whether $\mathrm{tp}_k$ is in the influence region of an active stop signal at \todo{time step}~$t'$.

Then, we define an intent indicator $\Phi_{\mathrm{intent}}(\mathrm{tp}_i, \mathrm{tp}_j)$ that~\todo{captures intent conflicts between participants $\mathrm{tp}_i, \mathrm{tp}_j \in \mathcal{TP}^t$,~formalized as spatial overlap of their future trajectories} by Eq.~\ref{eq:phi-intent},
\begin{equation}\label{eq:phi-intent}
\small
\Phi_{\mathrm{intent}}(\mathrm{tp}_i, \mathrm{tp}_j) = \mathds{1}\!\left[\mathrm{IS}(\Omega_i^t,\, \Omega_j^t)\right],
\end{equation}
where $\Omega_i^t = \bigcup_{t' \in \{t,\ldots,t+H\}} \bar{B}_i^{t'}$ denotes the trajectory corridor of $\mathrm{tp}_i$, \ie the union of the predicted bounding boxes over the assessment horizon~\cite{dresner2008multiagent}.
In practice, when~$\mathrm{tp}_i$~is a stalled AV with stationary predicted future trajectory $\bar{\tau}_i^t$, naively building $\Omega_i^t$ from $\bar{\tau}_i^t$ would collapse the corridor to $\mathrm{tp}_i$'s current location and lose its forward intent. We therefore populate $\Omega_i^t$ with $\mathrm{tp}_i$'s trajectory corridor computed just before the stall, so that $\Phi_{\mathrm{intent}}$ can still flag intent conflicts involving $\mathrm{tp}_i$.

\todo{Combining the stalling indicator and the intent indicator, we define the wait-for relation $\mathcal{E}^t$ by Eq.~\ref{eq:phi-wait},}
\begin{equation}\label{eq:phi-wait}
\small
\mathcal{E}^t = \big\{(\mathrm{tp}_i, \mathrm{tp}_j) \in \mathcal{TP}^t \times \mathcal{TP}^t \;\big|\; \mathrm{Stall}_i^t\,\land\, \Phi_{\mathrm{intent}}(\mathrm{tp}_i, \mathrm{tp}_j) \big\}.
\end{equation}
where an edge $(\mathrm{tp}_i, \mathrm{tp}_j) \in \mathcal{E}^t$ indicates that the stalled $\mathrm{tp}_i$~is blocked by $\mathrm{tp}_j$'s conflicting trajectory corridor. Finally, we construct the wait-for graph $\mathcal{G}^t = (\mathcal{TP}^t,$ $\mathcal{E}^t)$, and define the deadlock-unsafe AV set $\mathcal{U}^t_{\mathrm{dead}}$ by Eq.~\ref{eq:deadlock-hazard},
\begin{equation}\label{eq:deadlock-hazard}
\small
\mathcal{U}^t_{\mathrm{dead}} = \big\{\, \mathrm{av}_i \in \mathcal{V}_{\mathrm{AV}}^t \;\big|\; \mathrm{av}_i \in \mathrm{Cycles}(\mathcal{G}^t) \,\big\},
\end{equation}
where $\mathrm{Cycles}(\mathcal{G}^t)$ represents the set of AVs lying on directed cycles in $\mathcal{G}^t$, identified via depth-first search. \CR{Fig.~\ref{img:motivation:b} shows an example of a deadlock, in which AV1 and AV2 both stall at the unsignalized intersection, and their trajectory corridors computed before the stall overlap. Hence, $\mathcal{E}^t$ contains edges in both directions between them, which form a directed cycle, and both AVs are flagged as deadlock-unsafe.}

\textbf{Stalling Hazard.}
A stalling hazard arises when an AV~remains persistently stationary without any legitimate cause, \eg failing to resume after a green light or blocked~by~spurious~obstacles. We define the stalling-unsafe AV set $\mathcal{U}^t_{\mathrm{stall}}$ by Eq.~\ref{eq:stalling-hazard},
\begin{equation}\label{eq:stalling-hazard}
    \small
\mathcal{U}^t_{\mathrm{stall}} = \big\{\mathrm{av}_i \in \mathcal{V}_{\mathrm{AV}}^t \big| \mathrm{Stall}_i^t \land \deg^{+}_{\mathcal{G}^t}(\mathrm{av}_i) = 0 \big\},
\end{equation}
where $\deg^{+}_{\mathcal{G}^t}(\mathrm{av}_i) = 0$ denotes that $\mathrm{av}_i$ has zero out-degree in the wait-for graph $\mathcal{G}^t$, \ie not waiting for any traffic participant.

Then, \tool  aggregates the flagged AVs from all four~oracles into the \todo{current-time-step} unsafe AV set $\mathcal{U}^t_{\mathrm{cur}}$.
As \todo{per-time-step} oracle outputs may flicker under perception and prediction noise, we apply a cooldown mechanism before producing the final unsafe AV set; \ie the cooldown set is defined by Eq.~\ref{eq:cooldown}.
\begin{equation}\label{eq:cooldown}
    \small
\mathcal{U}^t_{\mathrm{cool}} = \big\{\, \mathrm{av}_i \in \mathcal{V}_{\mathrm{AV}}^t \;\big|\; t - t^{\mathrm{last}}_i < w_{\mathrm{hold}} \,\big\},
\end{equation}
where $w_{\mathrm{hold}}$ is the cooldown window length; and $t^{\mathrm{last}}_i$ is the most recent \todo{time step} at which $\mathrm{av}_i$ was flagged as unsafe,~and $t^{\mathrm{last}}_i$ resets to $t$ whenever $\mathrm{av}_i \in \mathcal{U}^t_{\mathrm{cur}}$. Finally, the~hazard~assessment gets the unsafe AV set $\mathcal{U}^t = \mathcal{U}^t_{\mathrm{cur}} \cup \mathcal{U}^t_{\mathrm{cool}}$~at \todo{time~step}~$t$.

\begin{figure}[!t]
    \centering
    \includegraphics[width=.52\linewidth]{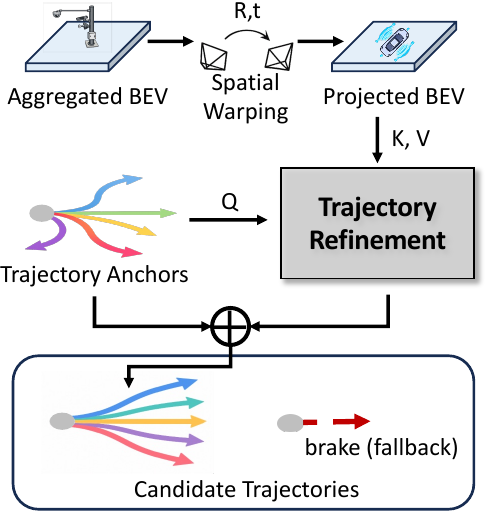}
    \caption{Architecture of Ego-level Candidate Generation}
    \label{fig:JHM}
\end{figure} 
\subsection{Joint Hazard Mitigation}
\label{sec:joint-mitigation}
To mitigate the hazards identified by the hazard assessment, \tool performs the joint hazard mitigation to produce a coordinated set of safe reference trajectories $\{\hat{\tau}_i^t\}_{\mathrm{av}_i \in \mathcal{U}^t}$.~It~consists of two stages, \ie \textit{ego-level candidate generation} first~generates a diverse pool of candidate trajectories for each unsafe AV from its ego-centric view, and then \textit{BnB-based~joint selection} performs scene-level coordination via branch-and-bound search to assign coordinated safe reference trajectories.

\textbf{Ego-level Candidate Generation.}
Inspired by recent end-to-end ADSes that leverage trajectory anchors to produce~multi-modal driving trajectories~\cite{jiang2024vadv2,li2025end}, we generate per-AV~candidate trajectories by refining a trajectory anchor library with BEV-conditioned residuals, providing diverse and realistic~override candidates. The anchor library $\{c^m\}_{m=1}^{M}$ \CR{contains $M$ representative trajectories, \ie the centroids obtained by} K-means clustering of expert trajectories from the \todo{Bench2Drive} training data~\cite{jia2024bench}, with each $c^m$ expressed in the \todo{ego coordinate frame} and covering common driving behaviors (\eg through, turn, lane change, and brake). 
As shown~in~Fig.~\ref{fig:JHM},~for~each~unsafe AV $\mathrm{av}_i$, we project $\mathrm{BEV}_{\mathrm{agg}}^t$ into its ego coordinate frame by the inverse pose $(T^t_{i \to \mathrm{RSU}})^{-1}$, yielding the projected BEV $\mathrm{BEV}_{\mathrm{agg} \to i}^t$ \todo{aligned with the anchor library}. Each anchor $c^m$~is then refined by a cross-attention residual head with~anchor~fea-tures as queries and $\mathrm{BEV}_{\mathrm{agg} \to i}^t$ as keys and values by Eq.~\ref{eq:residual},
\begin{equation}
\label{eq:residual}
\small
c_i^m = c^m + \mathrm{MLP} \Bigl(\mathrm{Attn}\bigl(\mathrm{TE}(c^m), \mathrm{BEV}_{\mathrm{agg} \to i}^t\bigr)\Bigr),
\end{equation}
where $\mathrm{TE}(\cdot)$ is an MLP-based trajectory encoder, and $\mathrm{Attn}(\cdot, \cdot)$ denotes a multi-head cross-attention operator. \CR{Intuitively, $\mathrm{Attn}$ retrieves scene context (\eg nearby obstacles) relevant to the anchor feature $\mathrm{TE}(c^m)$ from $\mathrm{BEV}_{\mathrm{agg} \to i}^t$, and $\mathrm{MLP}$ converts it into per-waypoint offsets that adapt $c^m$ into its refinement $c_i^m$ for $\mathrm{av}_i$.}

The final candidate set augments the $M$ \CR{refined} candidates $\{c_i^m\}_{m=1}^{M}$ with a deterministic brake trajectory by Eq.~\ref{eq:candset},
\begin{equation}
\label{eq:candset}
\small
\mathcal{C}_i \;=\; \bigl\{c_i^m\bigr\}_{m=1}^{M} \;\cup\; \bigl\{c_i^{\mathrm{brake}}\bigr\},
\end{equation}
where $c_i^{\mathrm{brake}}$ \CR{is a trajectory that decelerates at a constant rate from} $\mathrm{av}_i$'s current state, serving as a conservative fallback.

\todo{To preserve the behavioral diversity of the anchor library,~we train the residual head on Bench2Drive~\cite{jia2024bench} with the winner-take-all strategy~\cite{zhou2023query}, which supervises only the refined~trajectory from the anchor closest to each expert trajectory.}

\begin{algorithm}[!t]
    \caption{BnB-based Joint Selection}
    \footnotesize
    \label{alg:bnb}
    \LinesNumbered
    \KwIn{unsafe AV set $\mathcal{U}^t$, candidate trajectory sets $\{\mathcal{C}_i\}_{\mathrm{av}_i \in \mathcal{U}^t}$, \\
        \, \, \, \, \, \, auxiliary planning information $\{a_i^t\}_{\mathrm{av}_i \in \mathcal{U}^t}$, \\
        \, \, \, \, \, \, scene-level information $\mathcal{S}^t = (\{(\bar{s}_k^t, \bar{\tau}_k^t)\}, \mathcal{M}_{\mathrm{RSU}}, \mathcal{S}_{\mathrm{RSU}}^t)$, \\
        \, \, \, \, \, \, persistent yield memory $\mathcal{Y}$ from the previous time step}
    \KwOut{coordinated safe reference trajectories $\{\hat{\tau}_i^t\}_{\mathrm{av}_i \in \mathcal{U}^t}$,\\ 
    \, \, \, \, \, \, \, updated yield memory $\mathcal{Y}$}

    $\mathcal{O}^t \leftarrow \{\bar{\tau}_k^t \mid \mathrm{tp}_k \in \mathcal{TP}^t \setminus \mathcal{U}^t\}$\; \label{bg_derive}
    \ForEach{$\mathrm{av}_i \in \mathcal{U}^t$}{\label{pool_begin}
        $\mathcal{C}_i \leftarrow T^t_{i \to \mathrm{RSU}} \cdot \mathcal{C}_i$\;\label{ego2world}
        $\mathcal{C}_i \leftarrow$ HardPrune$\bigl(\mathcal{C}_i, \mathcal{O}^t, \mathcal{M}_{\mathrm{RSU}}\bigr)$\;\label{hard_prune}
        $\mathcal{J}_i \leftarrow \{c \mapsto J(c)\}_{c \in \mathcal{C}_i}$\;\label{cost_precompute}
        $\mathcal{C}_i \leftarrow$ DiverseTopK$\bigl(\mathcal{C}_i,\ \mathcal{J}_i\bigr)$\;\label{diverse_topk}
    }\label{pool_end}

    $\mathcal{Y} \leftarrow$ UpdateYieldRules$\bigl(\{\mathcal{C}_i\}_{\mathrm{av}_i \in \mathcal{U}^t},\, \{\mathcal{J}_i\}_{\mathrm{av}_i \in \mathcal{U}^t},\, \mathcal{Y}\bigr)$\;\label{yield_estab}
    $\mathcal{G}_{\mathrm{COL}}^t \leftarrow$ BuildConflictGraph$\bigl(\{\mathcal{C}_i\}_{\mathrm{av}_i \in \mathcal{U}^t},\, \mathcal{Y}\bigr)$\;\label{conflict_build}

    Sort $\mathcal{U}^t$ in ascending order of the size of candidate set $|\mathcal{C}_i|$\;\label{mrv}
    \textbf{global} $\mathbf{x}^\ast \leftarrow \{c_i^{\mathrm{brake}}\}_{\mathrm{av}_i \in \mathcal{U}^t}$, $J^\ast \leftarrow \sum_{\mathrm{av}_i \in \mathcal{U}^t} \mathcal{J}_i(c_i^{\mathrm{brake}})$\;\label{init_best}
    BnB$(\emptyset,\ 0)$; $\{\hat{\tau}_i^t\}_{\mathrm{av}_i \in \mathcal{U}^t} \leftarrow \mathbf{x}^\ast$ \; \label{branch_call}
    \Return $\{\hat{\tau}_i^t\}_{\mathrm{av}_i \in \mathcal{U}^t},\ \mathcal{Y}$\;

    \BlankLine
    \SetKwProg{Fn}{Procedure}{:}{}
    \Fn{BnB$(\mathbf{x},\ \mathit{lb})$}{
        \lIf{$\mathit{lb} \geq J^\ast$}{\Return\label{cost_prune}}
        \If{$|\mathbf{x}| = |\mathcal{U}^t|$}{
            $J^\ast \leftarrow \mathit{lb}$;\ \ $\mathbf{x}^\ast \leftarrow \mathbf{x}$;\ \ \Return\;\label{leaf}
        }
        $\mathrm{av}_i \leftarrow$ next unassigned AV from sorted $\mathcal{U}^t$\;\label{branch_pick}
        \ForEach{$c \in \mathcal{C}_i$ in ascending $\mathcal{J}_i(c)$}{\label{order_cand}
            \lIf{$\exists\,(j,c') \in \mathbf{x}: \bigl((i,c),(j,c')\bigr) \in \mathcal{G}_{\mathrm{COL}}^t$}{\textbf{continue}\label{conf_prune}}
            BnB$\bigl(\mathbf{x} \cup \{i \mapsto c\},\ \mathit{lb} + \mathcal{J}_i(c)\bigr)$\;\label{recurse}
        }
    }
\end{algorithm}

\textbf{BnB-based Joint Selection.}
Given the unsafe AV set $\mathcal{U}^t$ with candidate trajectory sets $\{\mathcal{C}_i\}_{\mathrm{av}_i \in \mathcal{U}^t}$, \todo{the auxiliary planning information $\{a_i^t\}_{\mathrm{av}_i \in \mathcal{U}^t}$}, the scene-level information $\mathcal{S}^t = (\{(\bar{s}_k^t,\bar{\tau}_k^t)\},\mathcal{M}_{\mathrm{RSU}},\mathcal{S}_{\mathrm{RSU}}^t)$, and the persistent yield~memory~$\mathcal{Y}$ from the previous \todo{time step (\ie yield priorities between~unsafe AV pairs stored across time steps)}, we formulate joint mitigation as a discrete assignment problem over finite per-AV candidate pools.
We use branch-and-bound~(BnB)~to~select~one~safe~reference trajectory $\hat{\tau}_i^t \in \mathcal{C}_i$ per unsafe AV, which resolves all spatial conflicts while staying close to each ADS's auxiliary planning information $a_i^t$, alongside an updated $\mathcal{Y}$ for subsequent \todo{time steps}. Algorithm~\ref{alg:bnb} illustrates the joint selection process.

We first collect the non-controllable background trajectories $\mathcal{O}^t$ (Line~\ref{bg_derive}), which include both AVs deemed safe and \todo{other traffic participants} in the current time step, acting~as~fixed~obstacles for the collision check.
For each unsafe AV $\mathrm{av}_i$,~the candidates in $\mathcal{C}_i$ are then transformed from $\mathrm{av}_i$'s ego coordinate frame into the \todo{RSU coordinate frame} via $T^t_{i \to \mathrm{RSU}}$ (Line~\ref{ego2world}). Then, we prune three classes of candidates (Line~\ref{hard_prune}), \ie~(i)~candidates whose net displacement falls below a small motion threshold\todo{, leaving $c_i^{\mathrm{brake}}$ as the unique stop candidate};
(ii) candidates~that~would collide with any $\bar{\tau}_k^t \in \mathcal{O}^t$ within the horizon; and (iii) candidates that leave the drivable region obtained from $\mathcal{M}_{\mathrm{RSU}}$. The brake candidate $c_i^{\mathrm{brake}}$ is unconditionally retained, guaranteeing~a conservative~fallback. \todo{This pruning excludes~candidates that~are infeasible against the fixed environment (\ie background~trajectories $\mathcal{O}^t$ and the drivable region), and thus our joint search only needs to coordinate mutual conflicts among unsafe AVs.}

Next, we precompute a per-AV cost map $\mathcal{J}_i: \mathcal{C}_i \mapsto \mathbb{R}$ that scores every surviving candidate (Line~\ref{cost_precompute}). The cost balances three concerns, as defined by Eq.~\ref{eq:cost},
\begin{equation}
\label{eq:cost}
\small
J(c) \;=\; J_{\mathrm{ref}}(c) \;+\; J_{\mathrm{bg}}(c) \;+\; J_{\mathrm{AV}}(c),
\end{equation}
where each term is non-negative and bounded to a comparable scale near $[0,1]$.
$J_{\mathrm{ref}}$ penalizes the candidate's mean deviation from a reference trajectory realized from $a_i^t$~and~$\mathcal{M}_{\mathrm{RSU}}$,~favoring candidates that stay close to the ADS's auxiliary~planning~information. $J_{\mathrm{bg}}$ decays exponentially with the candidate's~\CR{minimum} distance to $\mathcal{O}^t$ over the horizon\CR{, measured between time-aligned waypoints}, rewarding candidates~that keep a safe distance from background traffic. $J_{\mathrm{AV}}$ decays exponentially with the \CR{distance from the candidate's waypoints to the current positions of other unsafe AVs, since their trajectories are yet to be selected}, biasing the search toward likely-feasible regions first.
The brake candidate $c_i^{\mathrm{brake}}$ is assigned a fixed cost larger than $J(c)$ for any other $c \in \mathcal{C}_i$, so ascending-cost ordering of $\mathcal{C}_i$ always exhausts the remaining candidates before falling back to brake. Since $J(c)$ is a pure function of $c$ and $\mathcal{S}^t$, $\mathcal{J}_i$ is precomputed once per \todo{time step}, and reused as $O(1)$ lookups throughout our BnB search, keeping the inner loop free of per-node cost evaluation. Finally, $\mathcal{C}_i$ is reduced to \todo{$\kappa$ low-cost candidates covering distinct behavioral modes} (Line~\ref{diverse_topk}), bounding the BnB search width.

After per-AV candidate pool construction, we update the~persistent yield memory $\mathcal{Y}$ for the current \todo{time step} (Line~\ref{yield_estab}). \todo{$\mathcal{Y}$ persists yield priorities across time steps so that once a yield priority is established between two AVs, it remains consistent in subsequent steps, avoiding oscillation.} For each unsafe AV pair $\mathrm{av}_i, \mathrm{av}_j \in \mathcal{U}^t$ without a yield priority already registered in $\mathcal{Y}$, we take each AV's lowest-cost candidate as its representative trajectory, and check whether the corresponding corridors intersect. If the corridors do not intersect, no rule is established. Otherwise, the AV whose representative trajectory enters the intersection of the two corridors earlier is granted priority, and the other AV is recorded as yielding to it. 

Based on the updated yield memory, we precompute~the~conflict graph $\mathcal{G}_{\mathrm{COL}}^t$ over candidate pairs $(c, c') \in \mathcal{C}_i \times \mathcal{C}_j$~for~distinct unsafe AVs (Line~\ref{conflict_build}). Specifically, for each pair $(c, c')$, we first check whether their trajectory corridors intersect.
If their corridors intersect and the two candidates spatially~overlap~at any \todo{time step} $t' \in \{t,\ldots,t+H\}$, the pair is declared~as~conflicting.
When the corridors intersect without same-time overlap, the later-arriving AV is the implied yielder, and the pair~is~declared as conflicting if this relation contradicts a rule in $\mathcal{Y}$. Although the precomputed lookup tables $\{\mathcal{J}_i\}$ and $\mathcal{G}_{\mathrm{COL}}^t$ may hold more entries than the BnB ultimately visits, the per-entry computation is independent and can be batched in parallel upfront, making it faster than evaluating each entry on demand during the search.

Equipped with $\{\mathcal{J}_i\}$ and $\mathcal{G}_{\mathrm{COL}}^t$, we run the BnB search to assemble $\{\hat{\tau}_i^t\}_{\mathrm{av}_i \in \mathcal{U}^t}$. \CR{A joint assignment $\mathbf{x}$ maps each unsafe AV $\mathrm{av}_i$ to one candidate in $\mathcal{C}_i$, and its cost $J(\mathbf{x})$ is the sum of $\mathcal{J}_i$ over the assigned candidates.} BnB search systematically explores joint assignments while pruning subtrees~whose lower-bound cost exceeds the cost of the current best feasible solution, which means they cannot improve on the current~solution. The unsafe AVs are sorted by ascending $|\mathcal{C}_i|$, and thus the most-constrained AV is branched first (Line~\ref{mrv}). The current best solution $(\mathbf{x}^\ast, J^\ast)$ is initialized to the all-brake assignment (Line~\ref{init_best}), providing a guaranteed feasible fallback whose cost serves as the initial upper bound. The recursion (Lines~\ref{cost_prune}-\ref{recurse}) then prunes branches by the cost bound, expands candidates in ascending $\mathcal{J}_i(c)$ order, and skips any candidate that conflicts with an already-assigned one in $\mathcal{G}_{\mathrm{COL}}^t$. Upon completion, $\mathbf{x}^\ast$~is written into $\{\hat{\tau}_i^t\}$ (Line~\ref{branch_call}).

Finally, the joint hazard mitigation sends the safe reference trajectories $\{\hat{\tau}_i^t\}_{\mathrm{av}_i \in \mathcal{U}^t}$ to the corresponding unsafe AVs~to~resolve the hazards identified by the hazard assessment.

\subsection{\todo{Control Switch}}
\label{sec:control-override}

To allow the safe reference trajectories $\{\hat{\tau}_i^t\}_{\mathrm{av}_i \in \mathcal{U}^t}$ to take effect on unsafe AVs, each participating AV runs~a~control~switch module that is conditionally activated upon receiving~its~assigned safe reference trajectory $\hat{\tau}_i^t$ from~the~RSU. Specifically, when no $\hat{\tau}_i^t$ is received, the AV's original ADS-generated \CR{planned}~trajectory $\tau_i^t$ is passed through to the vehicle controller.
Otherwise, the control switch uses $\hat{\tau}_i^t$ as the trajectory finally executed~by the vehicle, and tracks it through our \todo{\textit{reference-guided override}, implemented as a model predictive~controller~(MPC) that~minimizes the tracking error to $\hat{\tau}_i^t$ over the MPC horizon under a kinematic bicycle model~\cite{ge2021numerically}, with additional penalties on control magnitude~and~rate~\cite{liu2025udmc}}.

The core of the control switch is the pass-through-or-override logic conditioned on $\hat{\tau}_i^t$, while the inner tracker is interchangeable. The MPC above is only a reference implementation, and any vendor-supplied controller (\eg PID, or learned policy) can substitute it without altering the RSU-side pipeline, preserving \tool's modular and ADS-agnostic~design.

\section{Evaluation}
We have implemented a prototype of \tool with \todo{7,959} lines of Python code. The source code and data of our work are available at~\cite{website1}.
To evaluate the effectiveness and efficiency of \tool, we design the following \CR{six} research questions.

\begin{itemize}[leftmargin=*]
    \item \textbf{RQ1 Effectiveness Evaluation}: How effective is \tool~in providing cooperative runtime safety assurance for various heterogeneous ADSes in mixed-traffic environments?
    \item \textbf{RQ2 Efficiency Evaluation}: Can \tool satisfy runtime latency constraints in mixed-traffic environments?
    \item \textbf{RQ3 Accuracy Evaluation}: How accurately can the hazard assessment identify unsafe AVs?
    \item \textbf{RQ4 Ablation Study}: How does each key module of \tool contribute to its overall effectiveness?
    \item \textbf{RQ5 Mixed-Deployment Evaluation}: Does \tool remain effective when coexisting AVs run different ADSes?
    \item \CR{\textbf{RQ6 V2X Robustness Evaluation}: How robust is \tool to V2X delay and packet loss?}
\end{itemize}

\subsection{Evaluation Setup}

\textbf{Benchmark.}
We adopt the InterDrive benchmark~\cite{liu2025colmdriver},~a~recent closed-loop benchmark built on the V2Xverse~\cite{liu2025towards}~platform that systematically assesses multi-vehicle interaction~in CARLA simulator~\cite{Dosovitskiy17}. InterDrive comprises 10 typical~multi-vehicle scenario types with reference to the pre-crash typology of NHTSA~\cite{NHTSA}, organized into three categories,~\ie InterDrive-IC where vehicles enter, meet, and exit an intersection from different directions, InterDrive-LM where vehicles merge into the same lane across parallel lanes, T-junctions, and highway ramps, and InterDrive-LC where vehicles traveling in parallel execute lane-change maneuvers that intersect adjacent vehicles' trajectories. We use the NPC-enriched configuration~\cite{liu2025colmdriver},~which yields 46 routes (\ie 18 for IC, 20 for LM, and 8 for~LC)~across 3 towns, each assigning 2 to 8 AVs on overlapping~paths~and populated with background vehicles, pedestrians, and cyclists. The driving task requires each AV to reach its destination while interacting safely with surrounding traffic~participants.

\textbf{Target ADSes.}
We integrate \tool with seven state-of-the-art ADSes spanning five families to evaluate its applicability across heterogeneous ADSes: (i) a modular ADS, Autoware~\cite{autoware}; (ii) three end-to-end ADSes, TCP~\cite{wu2022trajectoryguided}, UniAD~\cite{hu2023planning}, and VAD \cite{jiang2023vad}; (iii) an LLM-driven ADS, LMDrive~\cite{shao2024lmdrive}; (iv) a shared-perception cooperative ADS, CoDriving~\cite{liu2025towards}; and (v) an LLM-mediated negotiation cooperative ADS, CoLMDriver~\cite{liu2025colmdriver}.~For each AV, we run its native ADS with~the~publicly released official implementation, extracting the ADS-generated \CR{planned} trajectory and auxiliary planning information from~its~public~interfaces without modifying the underlying planning logic.
\todo{In our~prototype, all target ADSes share a single auxiliary~BEV~encoder and RSU-side fusion model, both pre-trained on DeepAccident~\cite{wang2024deepaccident}, instead of per-ADS BEV adapters~\cite{lu2024extensible}. This strategy keeps BEV-alignment quality constant across ADSes, and thus the~reported gains are attributable to \tool's safety assurance.}

\textbf{Baselines.}
We compare \tool against \textsc{Argus}~\cite{wang2025argus}, the state-of-the-art ego-centric runtime safety assurance framework for end-to-end ADSes, applying it to TCP, UniAD, and VAD within its design scope. To the best of our knowledge, no publicly available safety assurance framework applies to the other four target ADSes (\ie LMDrive, CoDriving, CoLMDriver, and Autoware). For all seven target ADSes, we additionally report their native ADS baselines without assurance as comparison.

\textbf{Metrics.}
To measure \tool's safety assurance effectiveness, we adopt two complementary metric groups, \ie~driving-performance metrics for the overall driving quality, and~violation counts for hazard prevention. Following \textsc{Argus}~\cite{wang2025argus},~we measure driving performance by four CARLA Leaderboard metrics~\cite{Dosovitskiy17}, \ie \textit{Route Completion} (\todo{$RC$, the completion~ratio of each route}), \textit{Infraction Score} (\todo{$IS$, a penalty factor starting at 1.0 and decreasing with each violation}),~\textit{Driving~Score}~($DS = RC \cdot IS$), and \textit{Success Rate} ($SR$, the fraction of routes~completed with a full-mark $DS$). Within each route, we aggregate $RC$ as the average, and $IS$ as the product across all AVs.
The final reported $RC$, $IS$, and $DS$ are then averaged across the 46 routes. For hazard prevention, we count four types of safety violations corresponding to the four hazard oracles, \ie vehicle collision ($Coll$), failure to obey stop signals ($Stop$), multi-vehicle deadlock ($Deadlock$), and vehicle~stalling~($Stall$).

\begin{table*}[!t]
    \caption{Results of the Driving Performance Improvement}
    \vspace{-5pt}
    \label{rq1:table1}
    \centering
    \begin{adjustbox}{width=0.98\linewidth}
    \begin{tabular}{lcccccccccccccccc}
        \toprule
        \multirow{2}{*}{\textbf{Approach}} & \multicolumn{4}{c}{\textbf{InterDrive-Total}} & \multicolumn{4}{c}{\textbf{InterDrive-IC}} & \multicolumn{4}{c}{\textbf{InterDrive-LM}} & \multicolumn{4}{c}{\textbf{InterDrive-LC}}\\
        \cmidrule(lr){2-5} \cmidrule(lr){6-9} \cmidrule(lr){10-13} \cmidrule(lr){14-17}
         & $DS\uparrow$ & $RC\uparrow$ & $IS\uparrow$ & $SR\uparrow$ & $DS\uparrow$ & $RC\uparrow$ & $IS\uparrow$ & $SR\uparrow$ & $DS\uparrow$ & $RC\uparrow$ & $IS\uparrow$ & $SR\uparrow$ & $DS\uparrow$ & $RC\uparrow$ & $IS\uparrow$ & $SR\uparrow$  \\
        \midrule
        \textbf{TCP}
        & 71.35 & 89.41 & 79.23 & 50.00
        & 69.25 & 78.02 & 85.83 & 44.44
        & 84.12 & 96.12 & 87.04 & 65.00
        & 44.12 & 98.29 & 44.83 & 25.00 \\
        \textbf{TCP-\textsc{Argus}}
        & 82.48 & 91.41 & 87.83 & 67.39
        & 81.54 & 89.28 & 88.63 & 61.11
        & 90.11 & 97.43 & 91.85 & 80.00
        & 65.51 & 81.14 & 76.00 & 50.00 \\
        \rowcolor[gray]{0.9} \textbf{TCP-\tool}
        & \textbf{92.42} & \textbf{94.22} & \textbf{97.22} & \textbf{82.61}
        & \textbf{89.52} & \textbf{93.07} & \textbf{96.44} & \textbf{72.22}
        & \textbf{97.59} & \textbf{97.97} & \textbf{99.63} & \textbf{85.00}
        & \textbf{100.00} & \textbf{100.00} & \textbf{100.00} & \textbf{100.00} \\
        \midrule

        \textbf{UniAD}
        & 35.19 & 87.59 & 36.72 & 13.04
        & 31.18 & \textbf{86.21} & 32.73 & 11.11
        & 50.94 & 91.63 & 52.89 & 20.00
        & 4.81 & 80.58 & 5.25 & 0.00 \\
        \textbf{UniAD-\textsc{Argus}}
        & 57.43 & 76.60 & 72.70 & 32.61
        & 64.03 & 75.93 & 77.81 & 38.89
        & 65.60 & 83.96 & 76.09 & 40.00
        & 22.17 & 59.69 & 52.73 & 0.00 \\
        \rowcolor[gray]{0.9} \textbf{UniAD-\tool}
        & \textbf{79.34} & \textbf{91.93} & \textbf{85.49} & \textbf{54.35}
        & \textbf{75.52} & 83.72 & \textbf{90.19} & \textbf{44.44}
        & \textbf{89.22} & \textbf{99.88} & \textbf{89.33} & \textbf{70.00}
        & \textbf{63.26} & \textbf{90.53} & \textbf{65.32} & \textbf{37.50} \\
        \midrule

        \textbf{VAD}
        & 30.85 & \textbf{96.81} & 31.13 & 13.04
        & 27.42 & \textbf{98.12} & 27.45 & 5.56
        & 45.44 & \textbf{97.84} & 45.77 & 25.00
        & 2.11 & \textbf{91.31} & 2.80 & 0.00 \\
        \textbf{VAD-\textsc{Argus}}
        & 57.02 & 79.30 & 67.85 & 28.26
        & 54.31 & 74.42 & 66.43 & 22.22
        & 69.56 & 92.21 & 74.06 & 40.00
        & 31.74 & 57.99 & 55.53 & 12.50 \\
         \rowcolor[gray]{0.9} \textbf{VAD-\tool}
        & \textbf{81.60} & 91.31 & \textbf{88.12} & \textbf{56.52}
        & \textbf{78.53} & 85.49 & \textbf{89.46} & \textbf{50.00}
        & \textbf{85.28} & 97.57 & \textbf{85.98} & \textbf{65.00}
        & \textbf{79.30} & 88.76 & \textbf{90.48} & \textbf{50.00} \\
        \midrule

        \textbf{LMDrive}
        & 64.47 & 74.45 & 88.24 & 26.09
        & 54.33 & 58.29 & 94.35 & 11.11
        & 73.23 & 85.29 & 86.15 & 35.00
        & 65.39 & 83.69 & 79.71 & 37.50 \\
        \rowcolor[gray]{0.9} \textbf{LMDrive-\tool}
        & \textbf{88.33} & \textbf{91.62} & \textbf{96.35} & \textbf{67.39}
        & \textbf{82.12} & \textbf{86.01} & \textbf{95.34} & \textbf{61.11}
        & \textbf{92.25} & \textbf{93.56} & \textbf{98.69} & \textbf{65.00}
        & \textbf{92.52} & \textbf{99.38} & \textbf{92.78} & \textbf{87.50} \\
        \midrule

        \textbf{CoDriving}
        & 68.04 & \textbf{95.34} & 70.85 & 52.17
        & 54.35 & \textbf{91.60} & 58.05 & 38.89
        & 93.33 & 96.83 & 96.49 & 75.00
        & 35.59 & \textbf{100.00} & 35.59 & 25.00 \\
        \rowcolor[gray]{0.9} \textbf{CoDriving-\tool}
        & \textbf{91.79} & 93.41 & \textbf{98.07} & \textbf{78.26}
        & \textbf{79.74} & 83.15 & \textbf{95.77} & \textbf{55.56}
        & \textbf{99.36} & \textbf{100.00} & \textbf{99.36} & \textbf{90.00}
        & \textbf{100.00} & \textbf{100.00} & \textbf{100.00} & \textbf{100.00} \\
        \midrule

        \textbf{CoLMDriver}
        & 79.83 & 92.43 & 82.64 & 71.74
        & 81.06 & \textbf{85.39} & 86.97 & \textbf{72.22}
        & 93.32 & \textbf{100.00} & 93.32 & \textbf{90.00}
        & 43.36 & 89.34 & 46.24 & 25.00 \\
        \rowcolor[gray]{0.9} \textbf{CoLMDriver-\tool}
        & \textbf{91.16} & \textbf{93.76} & \textbf{97.18} & \textbf{80.43}
        & \textbf{85.09} & 85.09 & \textbf{100.00} & \textbf{72.22}
        & \textbf{98.43} & \textbf{100.00} & \textbf{98.43} & \textbf{90.00}
        & \textbf{86.66} & \textbf{97.67} & \textbf{87.71} & \textbf{75.00} \\
        \midrule

        \textbf{Autoware}
        & 42.55 & 44.64 & 96.18 & 15.62
        & 40.86 & 44.90 & 90.05 & 13.33
        & 49.20 & 49.67 & 99.53 & 23.08
        & 27.31 & 27.31 & \textbf{100.00} & 0.00 \\
        \rowcolor[gray]{0.9} \textbf{Autoware-\tool}
        & \textbf{78.56} & \textbf{82.39} & \textbf{97.32} & \textbf{46.88}
        & \textbf{65.97} & \textbf{73.26} & \textbf{94.40} & \textbf{31.25}
        & \textbf{97.17} & \textbf{97.68} & \textbf{99.56} & \textbf{66.67}
        & \textbf{73.04} & \textbf{73.04} & \textbf{100.00} & \textbf{50.00} \\
        \bottomrule
    \end{tabular}
\end{adjustbox}
\end{table*}

\begin{table}[!t]
    \caption{Results of the Safety Violation Reduction}
    \vspace{-5pt}
    \label{rq1:table2}
    \centering
    \begin{adjustbox}{width=0.85\linewidth}
    \begin{tabular}{lcccc}
        \toprule
        \textbf{Approach} & $Coll\downarrow$ & $Stop\downarrow$ & $Deadlock\downarrow$ & $Stall\downarrow$ \\
        \midrule
        \textbf{TCP} &  21 & 5 & 4 & 4 \\
        \textbf{TCP-\textsc{Argus}} & 10 & 2 & 6 & 5 \\
        \rowcolor[gray]{0.9} \textbf{TCP-\tool} & \textbf{2} & \textbf{0} & \textbf{1} & \textbf{0} \\
        \midrule
        \textbf{UniAD} & 142 & \textbf{0} & 9 & 8 \\
        \textbf{UniAD-\textsc{Argus}} & 25 & \textbf{0} & 18 & 4 \\
        \rowcolor[gray]{0.9} \textbf{UniAD-\tool} & \textbf{12} & \textbf{0} & \textbf{2} & \textbf{1} \\
        \midrule
        \textbf{VAD} & 133 & 2 & \textbf{1} & 8 \\
        \textbf{VAD-\textsc{Argus}} & 26 & \textbf{0} & 16 & 5 \\
        \rowcolor[gray]{0.9} \textbf{VAD-\tool} & \textbf{6} & \textbf{0} & 5 & \textbf{0}  \\
        \midrule
        \textbf{LMDrive} & 15 & 3 & 10 & 24 \\
        \rowcolor[gray]{0.9} \textbf{LMDrive-\tool} & \textbf{3} & \textbf{0} & \textbf{3} & \textbf{8} \\
        \midrule
        \textbf{CoDriving} & 40 & \textbf{0} & 4 & 4\\
        \rowcolor[gray]{0.9} \textbf{CoDriving-\tool} & \textbf{1} & \textbf{0} & \textbf{3} & \textbf{1}  \\ 
        \midrule
        \textbf{CoLMDriver} & 18 & 2 & 7 & \textbf{0} \\
        \rowcolor[gray]{0.9} \textbf{CoLMDriver-\tool} & \textbf{6} & \textbf{0} & \textbf{1} & \textbf{0} \\
        \midrule
        \textbf{Autoware} &  \textbf{2} & \textbf{0} & 21 & 5\\
        \rowcolor[gray]{0.9} \textbf{Autoware-\tool} & \textbf{0} & \textbf{0} & \textbf{3} & \textbf{2} \\
        \bottomrule
    \end{tabular}
\end{adjustbox}
\end{table}

\textbf{RQ Setup.}
For \textbf{RQ1}, we compare \tool against the~baselines across the seven target ADSes separately, where each route deploys the same ADS for all AVs to isolate the per-ADS effect, and report the driving performance and violation counts.

For \textbf{RQ2}, we measure the \todo{per-time-step} wall-clock latency of each module of \tool over all routes, and report both the mean and P95 latency to characterize typical and tail behavior.

For \textbf{RQ3}, following \textsc{Argus}~\cite{wang2025argus}, we extract violation ground truth from the corresponding unprotected baseline runs using CARLA's infraction events. We treat an intervention as necessary if it occurs within 3 seconds prior to a violation.~Based on this, we count necessary interventions as true positives (TP), unnecessary interventions as false positives (FP), and violations not preceded by an intervention as false negatives (FN), and report $\text{precision} = \frac{\text{TP}}{\text{TP}+\text{FP}}$, $\text{recall} = \frac{\text{TP}}{\text{TP}+\text{FN}}$,~and~$\text{F}_{3}$~score~to~emphasize recall over precision in safety-critical settings.

For \textbf{RQ4}, we conduct module-level ablation studies~on~TCP as a representative ADS. We create \CR{five} variants~of~\tool, each replacing one key module with a simpler alternative,~\ie (i) \textit{w/o cooperative perception} disables V2X-based AV information sharing, and uses only RSU-side observations, (ii) \textit{w/o deadlock \& stall oracle} keeps only the collision and stop signal oracles in the hazard assessment, isolating the contribution of the interaction-induced oracles introduced by \tool, and (iii) \textit{w/o joint selection} replaces BnB-based joint selection with \CR{one of three simpler mechanisms, leaving other modules unchanged (\ie per-AV greedy selection, where each unsafe AV independently selects its best candidate; deterministic priority inspired by AIM~\cite{dresner2008multiagent}, where AVs take their best non-conflicting candidate in order of expected corridor entry, or brake; and randomized backoff, where each conflicting AV brakes for a random 1-8 time steps with a probability of 0.5).}

For \textbf{RQ5}, unlike \textbf{RQ1} where each route uses a single ADS type, the AVs within each route are assigned to TCP, LMDrive, and UniAD in rotation \CR{to simulate real-world roads with multiple vendors. These ADSes cover the end-to-end and LLM-driven families exhibiting different failure modes, and \tool reads only the V2X interface without identifying which ADS an AV runs.} The three ADSes appear in balanced proportions across the 46 routes (31.3\% TCP, 34.4\% LMDrive, and 34.4\% UniAD). The same AV-to-ADS mapping is used in both baseline and \tool runs to ensure a fair comparison.

\CR{For \textbf{RQ6}, to simulate imperfect V2X communication, we inject three perturbations (\ie 250 ms delay, 500 ms delay, and 30\% packet loss) into RSU-to-AV messages on TCP as a representative ADS, with all other modules unchanged. An AV that receives no safe reference trajectory executes its ADS-generated planned trajectory.}

\textbf{Environment.} We conduct all the experiments on Ubuntu 20.04.4 LTS servers with 4 NVIDIA GeForce RTX 3090 GPUs, Intel(R) Xeon(R) Silver 4310 @ 2.10GHz and 128GB memory. To mitigate the impact of randomness, all the experiments are conducted five times, and the average results are reported.

\subsection{Effectiveness Evaluation (RQ1)}

We evaluate the~effectiveness of \tool by reporting driving performance improvements and safety violation reductions.

\textbf{Driving Performance Improvement.}  As shown in Table~\ref{rq1:table1}, on InterDrive-Total, \tool achieves average improvements of 70.03\% in $DS$, 15.96\% in $RC$, 57.86\% in $IS$, and 162.29\% in $SR$ over the unprotected baselines. \CR{Among them, CoDriving and CoLMDriver gain 34.91\% and 14.19\% in $DS$, even though they already see beyond-line-of-sight hazards through V2X.} Across the three categories, \tool achieves average $DS$ improvements of 74.59\% on IC, 44.89\% on LM, and 784.27\% on LC, demonstrating that \tool's effectiveness generalizes across diverse interaction types, with the largest gains on~LC~where~the~unprotected~VAD and UniAD perform the worst. Compared to the ego-centric \textsc{Argus} baseline on TCP, UniAD, and VAD, \tool yields an average $DS$ gain of 31.10\%, confirming the advantage of scene-level cooperative intervention over ego-centric takeover.

\textbf{Violation Prevention.}
As shown in Table~\ref{rq1:table2}, \tool~substantially reduces four types of safety violations. On average across the seven target ADSes, \tool reduces $Coll$,~$Stop$, $Deadlock$ and $Stall$ by 91.91\%, 100.00\%, 67.86\% and~77.36\%, respectively. Overall, \tool reduces safety violations by 87.80\%. Compared to \textsc{Argus} on TCP, UniAD, and VAD, \tool also reduces safety violations, \todo{achieving a 75.21\% reduction.} Moreover, \textsc{Argus} inflates deadlocks on every~applicable ADS. We also conduct a breakdown analysis of the failure scenarios of each framework to understand their limitations.

\textbf{Breakdown of \textsc{Argus}.}
The residual failures of both \textsc{Argus} and \tool mainly stem from imperfect perception. Yet~\textsc{Argus} relies only on each AV's local perception, making it less effective against beyond-line-of-sight hazards. In addition, the increased deadlocks under \textsc{Argus} expose the limitation of ego-centric takeover, where AVs yield independently without scene-level coordination, producing locally reasonable yet jointly incompatible decisions. Through BnB-based joint selection and explicit yield-priority assignment, \tool avoids such states.

\textbf{Breakdown of \tool.} \tool does not fully eliminate $Deadlock$ and $Stall$ violations, most notably on VAD. These residual cases mainly arise when unsafe AVs are already too close for any forward-only candidate trajectories to resolve the conflict. Like most ADSes, \tool does not include reverse motion in its action space, and thus the BnB search falls back to a conservative all-stop assignment. This safety-oriented fallback often converts imminent collisions into less severe deadlock or stall states, thereby substantially reducing collision violations on VAD while leaving some residual liveness violations.

\subsection{Efficiency Evaluation (RQ2)}

We evaluate the runtime efficiency of \tool by profiling its \todo{per-time-step} wall-clock latency. The four modules' \todo{per-time-step} latencies are 27.55 / 52.30 ms (mean / P95)~for~the cooperative perception, 1.91 / 4.60 ms for the hazard assessment, 50.25 / 80.42 ms for the joint hazard mitigation (over active-intervention time steps), and below 0.01 ms for the~\todo{control switch}, with a mean total of 79.72 ms per active-intervention cycle. The logic-heavy kernels (\eg BnB-based joint selection and conflict-graph construction) are accelerated via Cython~\cite{behnel2010cython}.

To contextualize \tool's runtime overhead, we compare~it to the latency of seven state-of-the-art ADSes measured~in~our experiments. Across the ADSes, the mean \todo{per-time-step}~latency ranges from 168.6 ms (TCP, the fastest)~to~4,765.6~ms~(CoLMDriver, the slowest), all of which significantly exceed \tool's $\sim$80 ms budget. Since \tool mainly runs on the RSU side in parallel with each AV's native ADS, its overhead is fully subsumed by the ADS's own planning latency, and does not extend the critical path. Therefore, \tool delivers cooperative safety assurance while satisfying real-time constraints.

\subsection{Accuracy Evaluation (RQ3)}

\begin{table}[!t]
    \caption{Accuracy of the Hazard Assessment}
    \vspace{-5pt}
    \label{rq3:table1}
    \centering
    \begin{adjustbox}{width=0.75\linewidth}
    \begin{tabular}{lccc}
        \toprule
        \textbf{Approach} & Precision & Recall & $\text{F}_3$ \\
        \midrule
        \textbf{TCP-\textsc{Argus}} & 0.643 & 0.882 & 0.851 \\
        \rowcolor[gray]{0.9} \textbf{TCP-\tool} & 0.842 & 0.941 & 0.930 \\
        \midrule
        \textbf{UniAD-\textsc{Argus}} & 0.802 & 0.978 & 0.957 \\
        \rowcolor[gray]{0.9} \textbf{UniAD-\tool} & 0.902 & 0.982 & 0.973 \\
        \midrule
        \textbf{VAD-\textsc{Argus}} & 0.790 & 0.991 & 0.966 \\
        \rowcolor[gray]{0.9} \textbf{VAD-\tool} & 0.915 & 0.991 & 0.983 \\
        \midrule
        \rowcolor[gray]{0.9} \textbf{LMDrive-\tool} & 0.830 & 0.917 & 0.907 \\
        \rowcolor[gray]{0.9} \textbf{CoDriving-\tool} & 0.871 & 0.987 & 0.974 \\
        \rowcolor[gray]{0.9} \textbf{CoLMDriver-\tool} & 0.720 & 0.923 & 0.898 \\
        \rowcolor[gray]{0.9} \textbf{Autoware-\tool} & 0.774 & 0.857 & 0.848 \\
        \bottomrule
    \end{tabular}
\end{adjustbox}
\end{table}

We evaluate the accuracy of the hazard assessment~in~identifying unsafe AVs by reporting precision, recall, and $\text{F}_3$ score across the seven target ADSes. As shown in Table~\ref{rq3:table1}, \tool achieves an average precision of 0.836, recall of 0.943, and $\text{F}_3$ score of 0.930. Compared to \textsc{Argus} on TCP, UniAD, and VAD, \tool achieves average gains of 0.141 in precision, 0.021 in recall, and 0.037 in $\text{F}_3$ score. The high recall and $\text{F}_3$ score confirm that \tool's hazard assessment reliably identifies unsafe AVs in mixed-traffic environments.

\subsection{Ablation Study (RQ4)}

\begin{table*}[!t]
    \caption{Ablation Study for \tool on TCP}
    \vspace{-5pt}
    \label{rq4:table1}
    \centering
    \begin{adjustbox}{width=0.89\linewidth}
    \begin{tabular}{ccccccccccc}
        \toprule
        \multicolumn{3}{c}{\textbf{Step}} &  \multicolumn{4}{c}{\textbf{Driving Performance}} & \multicolumn{4}{c}{\textbf{Safety Violations}} \\
         \cmidrule(lr){1-3} \cmidrule(lr){4-7} \cmidrule(lr){8-11}
        Cooperative Perception & Deadlock \& Stall Oracle & Joint Selection & $DS\uparrow$ & $RC\uparrow$ & $IS\uparrow$ & $SR\uparrow$ & $Coll\downarrow$ & $Stop\downarrow$ & $Deadlock\downarrow$ & $Stall\downarrow$ \\
        \midrule
        \multicolumn{3}{c}{Original TCP} & 71.35 & 89.41 & 79.23 & 50.00 & 21 & 5 & 4 & 4 \\
        \ding{56} & \ding{51} & \ding{51} & 78.42 & 94.84 & 83.03 & 54.35 & 14 & 3 & 5 & 3 \\
        \ding{51} & \ding{56} & \ding{51} & 84.28 & 87.72 & 95.33 & 60.87 & \textbf{2} & \textbf{0} & 9 & 2 \\
        \midrule
        \ding{51} & \ding{51} & \CR{Per-AV Greedy} & 57.25 & \textbf{98.69} & 57.36 & 47.83 & 82 & \textbf{0} & \textbf{1} & \textbf{0} \\
        \CR{\ding{51}} & \CR{\ding{51}} & \CR{Deterministic Priority} & \CR{82.96} & \CR{98.41} & \CR{83.30} & \CR{76.09} & \CR{18} & \CR{\textbf{0}} & \CR{2} & \CR{2} \\
        \CR{\ding{51}} & \CR{\ding{51}} & \CR{Randomized Backoff} & \CR{71.78} & \CR{85.73} & \CR{75.25} & \CR{63.04} & \CR{26} & \CR{\textbf{0}} & \CR{8} & \CR{6} \\
        \midrule
        \rowcolor[gray]{0.9} \ding{51} & \ding{51} & \ding{51} & \textbf{92.42} & 94.22 & \textbf{97.22} & \textbf{82.61} & \textbf{2} & \textbf{0} & \textbf{1} & \textbf{0} \\
        \bottomrule
    \end{tabular}
\end{adjustbox}
\end{table*}

Table~\ref{rq4:table1} reports the ablation result on TCP.
Although the \textit{w/o cooperative perception} variant improves driving performance over the original TCP, its gains are substantially smaller than those of TCP-\tool, indicating that cooperative perception is essential for extending scene-level observability beyond~the RSU's local field of view. The \textit{w/o deadlock \& stall oracle} variant preserves TCP-\tool's reductions in $Coll$ and~$Stop$ violations, but increases $Deadlock$ beyond the original TCP baseline because collision-avoiding overrides may leave AVs in mutually blocked states that the disabled oracles no longer detect. \CR{With the same cooperative perception and hazard assessment, TCP-\tool still improves $DS$ over the \textit{w/o joint selection} variants (\ie by 61.43\% over per-AV greedy selection, 11.40\% over deterministic priority, and 28.75\% over randomized backoff), indicating that its gains come from scene-level coordination rather than perception alone. In particular, the per-AV greedy variant} performs worse than~the original TCP, with $Coll$ rising to nearly four times the baseline, because independently selected candidates can produce mutually inconsistent trajectories. All three modules contribute distinctly to \tool, and their combination delivers the effectiveness of cooperative safety assurance.

\subsection{Mixed-Deployment Evaluation (RQ5)}

\begin{table}[!t]
    \caption{Results of the Mixed-Deployment Evaluation}
    \vspace{-5pt}
    \label{rq5:table1}
    \centering
    \begin{adjustbox}{width=0.99\linewidth}
    \begin{tabular}{lcccccccc}
        \toprule
        \multirow{2}{*}{\textbf{Approach}} & \multicolumn{4}{c}{\textbf{Driving Performance}} & \multicolumn{4}{c}{\textbf{Safety Violations}} \\
        \cmidrule(lr){2-5} \cmidrule(lr){6-9}
         & $DS\uparrow$ & $RC\uparrow$ & $IS\uparrow$ & $SR\uparrow$ & $Coll\downarrow$ & $Stop\downarrow$ & $Deadlock\downarrow$ & $Stall\downarrow$  \\
        \midrule
        \CR{Mixed ADSes w/o \tool} & 55.16 & 88.21 & 61.39 & 30.43 & 51 & 4 & 6 & 18\\
        \rowcolor[gray]{0.9} \CR{Mixed ADSes w/ \tool} & \textbf{83.00} & \textbf{91.06} & \textbf{90.20} & \textbf{52.17} & \textbf{9} & \textbf{0} & \textbf{2} & \textbf{6}\\
        \bottomrule
    \end{tabular}
\end{adjustbox}
\end{table}

As shown in Table~\ref{rq5:table1}, in mixed deployment,~\tool~improves driving performance over the baseline without runtime assurance by 50.47\% in $DS$, 3.23\% in $RC$, 46.93\% in $IS$,~and 71.44\% in $SR$. It also reduces $Coll$, $Stop$, $Deadlock$, and $Stall$ violations by 82.35\%, 100.00\%, 66.67\%, and 66.67\%, respectively. These gains are comparable to the results in \textbf{RQ1}, confirming that \tool delivers cooperative runtime safety assurance even when coexisting AVs run different~ADSes.

\subsection{\CR{V2X Robustness Evaluation (RQ6)}}

\begin{table}[!t]
    \caption{\CR{Results of the V2X Robustness Evaluation on TCP}}
    \vspace{-5pt}
    \label{rq6:table1}
    \centering
    \begin{adjustbox}{width=0.99\linewidth}
    \begin{tabular}{lcccccccc}
        \toprule
        \multirow{2}{*}{\textbf{\CR{Condition}}} & \multicolumn{4}{c}{\textbf{\CR{Driving Performance}}} & \multicolumn{4}{c}{\textbf{\CR{Safety Violations}}} \\
        \cmidrule(lr){2-5} \cmidrule(lr){6-9}
         & \CR{$DS\uparrow$} & \CR{$RC\uparrow$} & \CR{$IS\uparrow$} & \CR{$SR\uparrow$} & \CR{$Coll\downarrow$} & \CR{$Stop\downarrow$} & \CR{$Deadlock\downarrow$} & \CR{$Stall\downarrow$} \\
        \midrule
        \CR{Original TCP} & \CR{71.35} & \CR{89.41} & \CR{79.23} & \CR{50.00} & \CR{21} & \CR{5} & \CR{4} & \CR{4} \\
        \CR{250 ms Delay} & \CR{90.03} & \CR{93.45} & \CR{93.57} & \CR{76.09} & \CR{5} & \CR{\textbf{0}} & \CR{\textbf{1}} & \CR{2} \\
        \CR{500 ms Delay} & \CR{83.05} & \CR{87.69} & \CR{92.03} & \CR{67.39} & \CR{6} & \CR{1} & \CR{3} & \CR{2} \\
        \CR{30\% Packet Loss} & \CR{90.27} & \CR{93.00} & \CR{94.93} & \CR{76.09} & \CR{3} & \CR{1} & \CR{2} & \CR{2} \\
        \rowcolor[gray]{0.9} \CR{TCP-\tool} & \CR{\textbf{92.42}} & \CR{\textbf{94.22}} & \CR{\textbf{97.22}} & \CR{\textbf{82.61}} & \CR{\textbf{2}} & \CR{\textbf{0}} & \CR{\textbf{1}} & \CR{\textbf{0}} \\
        \bottomrule
    \end{tabular}
\end{adjustbox}
\end{table}

\CR{As shown in Table~\ref{rq6:table1}, \tool improves $DS$ over the original TCP by 26.18\% under 250 ms delay, 16.40\% under 500 ms delay, and 26.52\% under 30\% packet loss. Delay makes each safe reference trajectory reflect an older scene, and packet loss skips intervention steps. Although both weaken the benefit of \tool, it still reduces safety violations compared to the original TCP under all three conditions.}

\section{Threats to Validity}

First, the benchmark selection may pose a threat to external validity. To mitigate this threat, we evaluate \tool~on~InterDrive~\cite{liu2025colmdriver}, the only publicly available closed-loop benchmark for V2X-enabled cooperative driving in mixed traffic,~with~its 10 scenario types aligned with the NHTSA pre-crash typology \cite{NHTSA} for evaluation in real-world accident distributions. %
These NHTSA-aligned scenarios cover the major interaction patterns expected in real-world mixed traffic, providing evidence~that \tool's safety assurance is not tied to a specific scenario.

Second, the quality of cooperative perception, motion~prediction, and BEV fusion may affect internal validity. \CR{\tool handles perception uncertainty through conservative measures (\eg bounding-box inflation in hazard oracles, a cooldown window against temporal flickering, and per-step re-assessment that corrects mis-predictions at the next observation).} To isolate \tool's safety assurance contribution from ADS-specific BEV adapters, our prototype evaluates all target ADSes under a uniform RSU-side BEV encoder and fusion model pre-trained on DeepAccident~\cite{wang2024deepaccident}. The required interfaces~are~specified as data formats (\ie BEV features, vehicle states, ADS~intentions, HD maps, and infrastructure states) rather than fixed perception implementations, and are compatible with established V2X~cooperative perception~\cite{lu2024extensible,wang2024deepaccident,liu2025towards,yu2025end,cui2022coopernaut}. Thus, the applicability of \tool scales with future literature advances. %

Third, the diversity of driving hazards presents a potential threat, as our current implementation focuses on four~representative safety violations. Compared with prior ego-centric runtime assurance~\cite{sun2024redriver,wang2025argus}, \tool explicitly defines an interaction-induced deadlock oracle that is critical in mixed traffic~where multiple AVs may make locally reasonable yet jointly incompatible decisions. The modular design of \tool~allows additional hazard oracles to be incorporated without changing the pass-through-or-override abstraction,~\eg~map-rule violations such as illegal U-turns or restricted-lane entry. \CR{Specifically, a new oracle needs to (i) be computable from the scene-level information constructed by cooperative perception; (ii) output the AVs requiring intervention; (iii) fit the per-step hazard assessment budget; and (iv) be mitigable by trajectory-level override, with its check reused in candidate pruning if it also constrains candidate trajectories.} While the current oracles cover a wide range of common safety violations,~other hazards, \eg sensor hardware faults or actuator~failures, require separate fault models and are beyond the scope of this work.

Finally, the deployment assumption underlying \tool may affect real-world applicability. \tool assumes an RSU-assisted prescriptive cooperation setting, where participating AVs exchange state, intention, and scene-level information with the RSU and execute the returned safe reference trajectories, which is consistent with the prescriptive cooperation class in SAE J3216~\cite{automation2020taxonomy} and recent V2X-enabled cooperative driving platforms~\cite{liu2025towards,cao2024segment}. 
Unlike centralized multi-vehicle planning approaches~\cite{wang2025knowledge,liu2024improved,huang2023parallel,huang2023decentralized} that replace the native ADSes, \tool preserves each AV's original ADS in normal operation, and only performs trajectory-level override for AVs identified as unsafe, imposing lighter deployment requirements.
Currently, we focus on the assurance logic built on top of this cooperation model and assume timely V2X message delivery \CR{in the main experiments. To mitigate this threat, we evaluate \tool under imperfect V2X communication, where \tool still reduces safety violations compared to the native ADS.}
As V2X \CR{communication becomes more reliable and} prescriptive cooperation interfaces become increasingly standardized, the real-world applicability of \tool is expected to be further enhanced.

\section{Related Work}

\textbf{Safety Assurance for ADSes.}
It aims to keep AVs within acceptable safety bounds by continuously monitoring driving hazards and intervening before potential safety violations~occur \cite{chen2022runtime,peng2023rta,sun2024redriver,wang2025argus}. Prior studies mainly monitor driving hazards by detecting unexpected driving conditions, distributional shifts, low model confidence, or rule-based misbehavior ~\cite{hussain2022deepguard,stocco2022thirdeye,candela2023risk,grewal2024predicting,stocco2020misbehaviour,qian2025collision,yu2024online}. However, these works are not designed~to~detect interaction-induced hazards such as deadlock that arise in mixed traffic, and focus on hazard detection rather than mitigation. 

Furthermore, various fallback strategies have been proposed to keep ADSes within safety bounds~\cite{chen2022runtime,peng2023rta,lin2020reachflow,yu2019fallback,hussain2022deepguard,xue2018fallback}. Rule-based safety models like RSS define interpretable safety distances and response rules for automated driving~\cite{shalev2017formal}, while emergency braking systems trigger conservative responses~to avoid potential collisions~\cite{kusano2012safety,zhang2024dual}. Runtime assurance frameworks like Simplex-Drive~\cite{chen2022runtime} and RTA-IR~\cite{peng2023rta} adopt switching architectures that transfer control to a verified safety~controller upon predicted violations, and fallback~policies~\cite{xue2018fallback}~prescribe pull-over maneuvers in emergencies. However, they rely on predetermined fallback responses rather than synthesizing scene-aware safe trajectories that resolve specific hazards.
  
Recent software engineering work has further studied runtime enforcement for ADS trajectories.
REDriver~\cite{sun2024redriver}~and~FixDrive~\cite{sun2025fixdrive} modify ADS-generated \CR{planned} trajectories to enforce~specification compliance, while \textsc{Argus}~\cite{wang2025argus} formulates runtime assurance as a hazard-aware takeover-and-recovery process~for ADSes.
While they demonstrate the benefit of adding a runtime safety layer over ADSes, their interventions remain ego-centric, with each ADS-controlled vehicle monitored and overridden independently and interaction-induced hazards in mixed traffic left unresolved.
In contrast, \tool provides an ADS-agnostic cooperative runtime safety assurance framework that monitors scene-level hazards and jointly generates coordinated~safe~reference trajectories for multiple unsafe AVs in mixed traffic.

\textbf{Cooperative Autonomous Driving.}
It leverages V2X~communication to extend the perception, prediction, and coordination capability of individual AVs. Extensive work~studies cooperative perception %
to alleviate occlusion~\cite{xu2024v2x,liu2020when2com,xu2023cobevt,lu2024extensible,wang2024deepaccident,cui2022coopernaut}. End-to-end cooperative driving systems further integrate shared perception into closed-loop driving~\cite{liu2025towards,yu2025end,you2026v2x}.~Beyond~perception sharing, decentralized planning approaches let~each~AV cooperatively plan with neighbors %
\cite{zhao2023multi,chen2024communication,zhou2022cooperative}. LLM-mediated negotiation approaches such as CoLMDriver~\cite{liu2025colmdriver}~further help multiple AVs exchange intentions and reach agreements~\cite{fang2025towards,gao2025langcoop}. However, they mainly enhance the information available to each ADS, while the driving decisions are still produced~by the independent driving policies and may remain unsafe. %

Another line of work focuses on centralized multi-vehicle planning %
\cite{lu2025vi,liu2024improved,huang2023parallel,huang2023decentralized,cao2024segment}. 
They often assume controllable CAV fleets, homogeneous cooperation protocols, or direct~replacement of the native driving policy. In contrast, \tool is not a centralized driving planner, but an ADS-agnostic runtime safety assurance layer that selectively overrides only unsafe AVs while preserving each AV's native ADS in normal operation.

\section{Conclusions}
We propose and implement \tool, an ADS-agnostic~cooperative runtime safety assurance framework for mixed-traffic~autonomous driving. %
Our large-scale experiments~on~the InterDrive benchmark with seven state-of-the-art ADSes demonstrate the effectiveness, efficiency, and generality of \tool.

\section{Data Availability}
The source code and data of our work are available at~\cite{website1}.

\section*{Acknowledgment}
\CR{This work was supported by the National Natural Science Foundation of China (Grant No. 92582205).}

{
\footnotesize
\bibliographystyle{IEEEtranS}
\bibliography{IEEEabrv, src/reference}
}

\end{document}